\documentclass[acmlarge,screen]{acmart}

\usepackage{color}
\usepackage{colortbl}
\usepackage{makecell}
\usepackage{multirow}
\usepackage[english]{babel}
\usepackage[utf8]{inputenc}
\usepackage{algorithm}
\usepackage[noend]{algpseudocode}

\usepackage{xcolor}

\AtBeginDocument{%
  \providecommand\BibTeX{{%
    \normalfont B\kern-0.5em{\scshape i\kern-0.25em b}\kern-0.8em\TeX}}}

\usepackage{wrapfig}
\usepackage{enumitem}
\usepackage{caption}
\usepackage{subcaption}
\usepackage{todonotes}

\definecolor{LightCyan}{rgb}{0.88,1,1}
\newcommand{\name}{{\em FlowSense}}             
\newcommand{\names}{{\em FlowSense\ }}  
\newcommand\review[1]{\textcolor{black}{#1}}
\newcommand\finalrev[1]{\textcolor{black}{#1}}

\begin{document}

\setcopyright{acmlicensed}
\acmJournal{IMWUT}
\acmYear{2022} \acmVolume{6} \acmNumber{1} \acmArticle{5} \acmMonth{3} \acmPrice{15.00}\acmDOI{10.1145/3517258}

\title{FlowSense: Monitoring Airflow in Building Ventilation Systems Using  Audio Sensing}

\author{Bhawana Chhaglani} \orcid{0000-0002-4060-4883}
\affiliation{%
  \institution{University of Massachusetts Amherst}
  \country{USA}
  }
\email{bchhaglani@cs.umass.edu}

\author{Camellia Zakaria}
\orcid{0000-0003-4520-9783}
\affiliation{%
  \institution{University of Massachusetts Amherst}
  \country{USA}
  }
\email{nurcamellia@cs.umass.edu}

\author{Adam Lechowicz}
\orcid{0000-0002-7774-9939}
\affiliation{%
  \institution{University of Massachusetts Amherst}
  \country{USA}
  }
\email{alechowicz@umass.edu}

\author{Jeremy Gummeson}
\orcid{0000-0002-7468-0569}
\affiliation{%
  \institution{University of Massachusetts Amherst}
  \country{USA}
  }
\email{gummeson@cs.umass.edu}

\author{Prashant Shenoy}
\orcid{0000-0002-5435-1901}
\affiliation{%
  \institution{University of Massachusetts Amherst}
  \country{USA}
  }
\email{shenoy@cs.umass.edu}

\renewcommand{\shortauthors}{Chhaglani et al.}



\begin{abstract}
Proper indoor ventilation through buildings' heating, ventilation, and air conditioning (HVAC) systems has become an increasing public health concern that significantly impacts individuals' health and safety at home, work, and school. While much work has progressed in providing energy-efficient and user comfort for HVAC systems through IoT devices and mobile-sensing approaches, ventilation is an aspect that has received lesser attention despite its importance. With a motivation to monitor airflow from building ventilation systems through commodity sensing devices, we present \name, a machine learning-based algorithm to predict airflow rate from sensed audio data in indoor spaces. Our ML technique can predict the state of an air vent---whether it is on or off---as well as the rate of air flowing through active vents. By exploiting a low-pass filter to obtain low-frequency audio signals, we put together a  privacy-preserving pipeline that leverages a silence detection algorithm to only \review{sense} sounds of from HVAC air vents when no human speech is detected. We also propose the Minimum Persistent Sensing (MPS) as a post-processing algorithm to reduce interference from ambient noise, including ongoing human conversation, office machines, and traffic noises. Together, these techniques ensure user privacy and improve the robustness of \name. We validate our approach yielding over 90\% accuracy in predicting vent status and 0.96 MSE in predicting airflow rate when the device is placed within 2.25 meters away from an air vent. Our approach can be generalized to environments with similar vent dimensions and geometry outlets. Additionally, we demonstrate how our approach as a mobile audio-sensing platform is robust to smartphone models, distance, and orientation. Finally, we evaluate \names privacy-preserving pipeline through a user study and a Google Speech Recognition service, confirming that the audio signals we used as input data are inaudible and inconstructible. 
\end{abstract}

\maketitle

\section{Introduction}
\label{sec:introduction}

Humans spend over 90\% of their lifetime in indoor spaces such as homes, office buildings, and schools \cite{epa}. Consequently, designing buildings for energy efficiency \cite{lam2014occupant} and comfort \cite{francis2019occutherm} of occupants has been a long-standing goal in recent years. For example, advanced sensing and communication technologies have designed smart buildings with real-time monitoring and control capabilities to avoid energy waste by turning off lighting, and air conditioning in unoccupied areas of a building \cite{beltran2014optimal}. Similarly, they have been employed to improve comfort through proper lighting and by personalizing thermal comfort to occupant preferences \cite{rabbani2016spot,wei2021low}

Since a building's heating, ventilation, and air-conditioning (HVAC) systems consume over 50\% of its total energy usage \cite{enegygov}, many efforts have focused on HVAC efficiency and comfort to improve the heating and cooling aspects of the systems. However, HVAC systems also include a third component, ventilation, which has seen much less attention despite its importance. Healthy air hygiene through proper ventilation is essential for human health and comfort \cite{sundell2011ventilation}. Being in poorly ventilated spaces can cause significant harm to occupants by increasing the risk of spreading infectious diseases \cite{epa}. Since the coronavirus pandemic, improving indoor ventilation has gained renewed interest since it is a crucial component for resuming professional life in workplaces in a safe and healthy manner. Proper ventilation for removing viral loads in ambient indoor air has been recommended as a key safety measure by the World Health Organization (WHO), Centers for Disease Control and Prevention (CDC), and American Society of Heating, Refrigerating and Air-Conditioning Engineers (ASHRAE) \cite{world2021roadmap,centers2021ventilation, ashraeOpen}. 

The amount of ventilation needed in indoor spaces depends on the occupancy levels and user-performed activities. Broadly speaking, indoor air quality depends on the amount of CO\textsubscript{2}, indoor pollutants, allergens, and viral particles present in the air.  Higher occupancy or indoor activities such as cleaning and cooking fumes will increase the levels 
of CO\textsubscript{2}, dust, and pollutants in indoor air. Ventilation systems are designed to maintain indoor air quality in the presence of such activities. 
 There are two broad ways to measure the efficacy of the ventilation system in buildings. The first approach is to directly measure air quality using various sensors. For example, many Internet of Things (IoT) products are now available to measure CO\textsubscript{2} and volatile organic compound (VOC) particles \cite{chojer2020development}, allowing users to monitor their surroundings through smartphone apps. The second approach is to measure the airflow through ventilation systems. The rate of airflow indicates the volume of fresh or filtered air entering an indoor space. A building's HVAC system typically includes duct sensors to measure airflow, but this data is internal to the building management system (BMS) and available only to facility managers. Building occupants do not have visibility into such data. Further, BMS sensors may be miscalibrated and sensed data may have measurement errors. These challenges motivate \emph{the need to develop low-cost and non-intrusive sensing techniques to monitor a building's ventilation system and expose the data to occupants and facility managers}. \review{Such functionality can also contribute as a significant sub-system to a fully integrated, smart ventilation solution that provides precise and real-time ventilation monitoring; facilitating ventilation only where and when needed.}

In this paper, we present \name, a novel sensing approach for monitoring airflow in building ventilation systems using audio sensing. Our audio sensing approach is based on ``listening'' to the sounds of air traveling from HVAC systems through the duct and estimating the airflow rate from the sensed audio signal. We hypothesize that the faint sounds of airflow from vents can be discerned by audio sensors (microphones) and used to determine the airflow rate in ventilation systems. Such audio sensors can either be deployed as a low-cost fixed sensing infrastructure or deployed on smartphones carried by building occupants. \review{To our knowledge, this is the first work that measures the rate of airflow using audio-sensing on smartphones.}

Since smartphones are ubiquitous, using them as mobile sensors is a viable approach for building monitoring. Other recent approaches have proposed using smartphones as temperature sensors for monitoring heating and cooling in indoor spaces \cite{breda2019hot}. In our cases, fixed or mobile audio sensors can expose information about ventilation in various rooms to end-users. At the same time, it also serves as a second source of ventilation data to facility managers to augment data from BMS sensors. Doing so can promote user awareness of indoor air quality by answering questions such as ``does the conference room have adequate air ventilation during the meeting?'' or ``is the classroom adequately ventilated during a lecture?''

Our approach needs to address two challenges in order to be practical. First, it needs to be non-intrusive by suppressing human speech or sensitive sounds that leak user privacy during audio sensing. Second, it needs to be robust to interference from ambient noise in the environment that can potentially affect airflow sensing through vents. Our work addresses both challenges through a combination of sensor processing techniques for enhancing robustness and user privacy. In designing, implementing, and evaluating \name, our paper makes the following contributions:

\begin{enumerate}
    \item We present machine learning-based algorithms to predict airflow rate from sensed audio data in indoor spaces. Our ML techniques can predict the state of an air vent---whether it is on or off---as well as the rate of air flowing through active vents. We present two techniques, silence period detection and Minimum Persistent Sensing (MPS), to enhance our machine learning methods in order to suppress human speech in sensed audio and reduce interference from ambient noise. Together these techniques ensure user privacy and improve the robustness of \name. 
    \item We implement a complete prototype of our approach as a software system on two hardware platforms. We realize \names as a low-cost fixed audio sensing platform on Arduino micro-controller and as a smartphone application for background sensing of ambient noise using the smartphone microphone. \review{The novelty of this approach is that is capable of monitoring airflow in indoor environments using only the microphone of a smartphone or any similarly-equipped device.}
    \item We deploy our fixed and mobile sensing prototypes in two office buildings on our campus and gather audio data from various vents and rooms in real-world settings and actual occupants. We use this data to demonstrate the feasibility of our approach. We make our source code and datasets available to researchers in open source and data form. 
    \item We conduct an extensive experimental evaluation of \names using fixed and mobile sensing platforms. Our results show that \names can efficiently determine the state of the vent with 99\% accuracy and estimate the rate of airflow with 95\% accuracy and remains accurate even at distances of up to 2.25 meters from air vents. Further, our system is robust to ambient noise since the proposed enhancements of silence period detection and MPS increase the overall accuracy by 77\%.  Our results also characterize key user factors that impact the performance of our approach, including the placement and orientation of sensors or phones, different vent types, and smartphone hardware specifications.
    \item We conduct a user study of 12 participants to validate the privacy-preserving nature of our approach in suppressing private human speech. Our user study and the use of an AI speech recognition system show that \names can suppress all human speech and preserve user privacy with nearly 100\% effectiveness.  
\end{enumerate}

\section{Background and Motivation}
\label{sec:background}

This section provides background on building ventilation systems and sensing techniques and then motivates our audio sensing approach.
\newline

\noindent\textbf{Building HVAC and Management Systems.}
The mechanical systems in offices and commercial buildings are responsible for heating, cooling, and ventilation (HVAC). Modern HVAC systems include a sensing infrastructure to monitor indoor temperature, humidity, and occupancy in various zones and have programmatic actuation capabilities to control the amount of heating, cooling, and conditioned air delivered to the building \cite{kohposter}. The sensing and actuation capabilities are exposed to facility managers through Building Management Systems (BMS) \cite{lowry2002factors}. Today's BMS exposes fine-grain monitoring capabilities and provides significant configuration capabilities. Importantly, however, BMS does not expose any data to the end-users. 
\newline

\noindent\textbf{IoT-based Building Automation.}
Numerous Internet of Things (IoT) products have emerged for building automation, mostly targeting residential rather than office buildings. These IoT products enable fine-grain monitoring of the indoor environment, including temperature \cite{shafiq2019reusable}, humidity, carbon dioxide \cite{Netatmo, jiang2011maqs}, and particulate matter, and expose monitored data through a smartphone. Other products such as smart thermostats  \cite{Nest, ecobee} and smart switches \cite{wemo, touchie2018residential} also enable remote programmatic control using frameworks such as IFTTT \cite{ifttt} and Apple HomeKit \cite{homekit}. Unlike BMS systems that focus on facility managers, IoT-based building automation has focused on end-users and building occupants. 
\newline

\noindent\textbf{Ventilation and Occupant Health.}
A building ventilation system removes stale air from indoor spaces and delivers fresh, filtered air. Replacement of air is done by venting out (through exhaust vents) stale air from inside, filtering it to remove indoor pollutants, mixing it with fresh outside air, and then delivering this filtered fresh air back to those indoor spaces. 

Proper ventilation is crucial for maintaining indoor air quality and for the respiratory health of the occupants since it removes CO\textsubscript{2} and indoor pollutants such as dust, allergens, and VOC particles generated from indoor activities. Improper ventilation can harm health, including increased chances of respiratory illness such as asthma and spreading infectious diseases from rising viral loads in indoor spaces \cite{Walker2021, Tang2021, WHO2021}.  While ventilation can be achieved by opening windows or using room-level vents to circulate air, our work focuses on building HVAC systems used in office or commercial buildings to provide ventilation.

As noted earlier, indoor ventilation can be monitored by measuring indoor air quality such as CO\textsubscript{2} levels or VOC particles or by monitoring the rate of airflow through the ventilation system. BMS sensors can monitor airflow through ducts and vents using air flow meters, pressure sensors, and vane anemometers \cite{mcwilliams2002review}. While highly accurate, these sensors are typically hard-wired and require commissioning to install and calibrate sensors. Further, they are accessible only to facility managers and not to occupants. 
\newline

\noindent\textbf{Mobile Sensors.}
Since smartphones are ubiquitous, many efforts have been developed to exploit the array of sensors available in smartphones as mobile sensors to monitor the environment. While the use of smartphones to monitor users' health is increasingly common, recent efforts have used phones to monitor buildings. For example, the sensors to monitor smartphone battery temperature have been used to monitor indoor air temperatures for HVAC systems \cite{breda2019hot}. Phones and other mobile devices have also been used to monitor occupancy levels on various floors for smart HVAC control \cite{trivedi2017ischedule, erickson2010occupancy}. Our work leverages smartphones (and other low-cost sensors) to monitor airflow in building ventilation systems. \review{We envision such capabilities being integrated with smart ventilation solutions that optimizes indoor air quality and enables precise ventilation monitoring based on human occupancy in real-time.}

\subsection{Motivation}
\label{sec:motivation}
Our work focuses on \emph{audio sensing} to monitor airflow through building ventilation systems. Specifically, our approach involves listening to the sounds of air flowing through vents and using these audio signals to infer the rate of airflow. To do so, we can use a commodity microphone as an audio sensor, capture ambient noises, and extract the sound of airflow from overall ambient noises sensed in the environment. Consider the frequency spectrum of an audio clip recorded in an office room to understand why this is feasible. Typically, there will be four broad categories of sound that will be discernible to an audio sensor(microphone): (i) human speech, if there are occupants present in the surroundings, (ii) background noise from the indoor environment(e.g., from office equipment and non-speech human activities), (iii) background noise from the outdoor environment that is audible indoors (e.g., traffic noise and nature sounds such as dogs barking or birds chirping), (iv) noise from ventilation equipment due to airflow. 

\begin{figure}[ht!]
    \centering
    \includegraphics[width=\columnwidth]{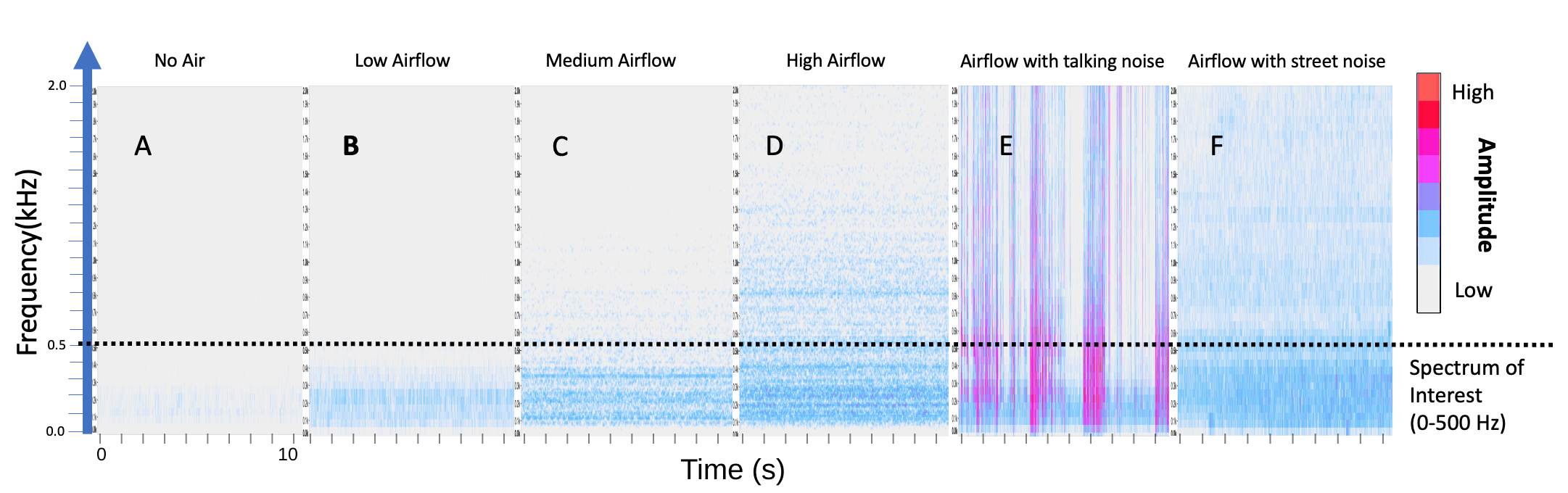}
    \vspace{-4mm}
    \caption{Audio frequency spectrum of airflow, human speech, and ambient noises.}
    \label{fig:intuition_spectrum}
\end{figure}

These sounds fall into different regions of the audio frequency spectrum. Typically, human speech lies in between 200 Hz to 5 kHz frequency bands, with much of the speech belonging to the medium frequency band of 400 Hz to 3 kHz range\cite{titze1998principles}.  Background noise from indoor activities and outdoor activities can belong to both medium and high-frequency bands (500Hz to 20KHz),  with some sounds belonging to low-frequency bands ($<$500Hz).
In contrast, airflow sounds through vents lie in low-frequency bands between 10 to 500Hz,  and some cases, belonging to mid-frequency bands up to 1000Hz. As shown in fig \ref{fig:intuition_spectrum}, low airflow rates range between 10-300 Hz, medium airflow rate lies between 10-400 Hz, while high airflow rates belong to 10Hz-1kHz. In all cases, the higher amplitudes are below 500 Hz. 

Figure \ref{fig:intuition_spectrum} shows multiple example audio frequency spectrums for various activities, \review{recorded using a laptop's integrated microphone. Note that while our motivation utilizes the laptop as a device, these insights are generalizable to other audio sensing modalities, as demonstrated in our experiments, which employ smartphone devices.} 
This example and our observations yield the following insights. Human speech, airflow, and ambient noises belong to different portions of the audio frequency spectrum, but there is also some overlap.  For example, much of human speech and ambient background noise lies in the mid and high-frequency bands (500Hz-2kHz), while airflow sounds from vents belong to low-frequency bands (10-500Hz). Hence, it should be feasible to extract the audio signal of airflow from other sounds (e.g., using a low pass filter). However, the figure also shows {\em there is also non-trivial interference between the various types of sounds.}  As shown in the region ``E'' of Figure \ref{fig:intuition_spectrum} -- airflow together with human speech, some low-frequency components of human speech depicted in pink interfere with noise from air vents in the 10-500 Hz band. The interference occurs when humans are speaking but not in silence periods between words and sentences. Similarly, region ``F'' of Figure \ref{fig:intuition_spectrum} shows that low-frequency components of background ambient noise can also interfere with noise from air vents. Thus, our audio sensing approach will need to be robust to interference between different types of sounds in an indoor environment. Further, any human speech in the low-frequency region(present after filtering out the mid and high-frequency components) should not leak any privacy. Our approach needs to address a second challenge through discernible words. 
\section{\names Design}

\begin{figure}[!h]
\centering
    \includegraphics[width=\linewidth]{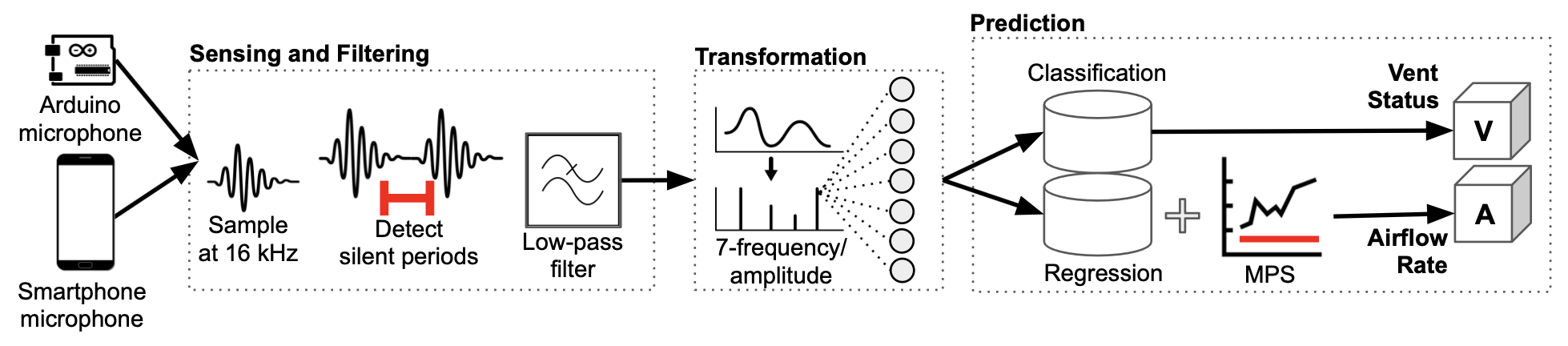}
    \caption{Overview of \name, consisting of three components.}
    \label{fig:sysOverview}
\end{figure}

In this section, we present the design of \name, our audio sensing approach for sensing airflow in building ventilation systems. The goal of \names is to predict whether the air vents are on/off in a room and estimate the airflow rate when the vent is on. We first present our machine learning techniques to determine the state of air vents and the rate of airflow. Next, we present two enhancements to our machine learning approach to ensure robustness to interference and preserve humans' privacy in the sensed data. Figure \ref{fig:sysOverview} depicts an overview of our \names approach, which we discuss in detail.

\subsection{Machine Learning Model}
\names uses machine learning models to estimate the state of an air vent and the rate of airflow through the vent. The problem of determining the vent state is a \emph{binary classification} problem where the audio signal is used to classify whether the vent is on or off. The related problem of estimating the rate of airflow is a \emph{regression} problem where the frequencies present in the signal and their amplitudes are used to determine the rate of air flowing through a vent.  As shown in Figure \ref{fig:sysOverview}, \names begins by capturing the audio signal from a microphone. The typical sampling frequency in commodity microphones is 16kHz. The captured audio signals will include low, medium, and high-frequency components. Since audio signals from air vents are predominantly in the low-frequency portion of the audio spectrum, the raw audio signal is sent through a low pass filter, which removes all medium and high-frequency components in the captured audio signal. In doing so, most of the human speech components in the signal are filtered out, and so are the medium and high-frequency portions of ambient background noise. 

In our current design, we use a low pass filter of 375 Hz since our experiments (see section \ref{sec:cut-off}) show that this provides good accuracy by retaining the most critical frequencies from the vent airflow noise. Further, this threshold is also a reasonable privacy filter since it removes most (but not all) frequency components of human speech, which primarily resides in 200 Hz to 5 kHz frequency bands \cite{titze1998principles}.

The transformed low-frequency audio signal is used to compute features that serve as inputs to our classification and regression models. We first apply the fast Fourier transform (FFT) to convert the audio signal from the time domain to the frequency domain. Each FFT uses a sample size of 256, including 16 ms of audio data, and transforms the time domain data into seven frequencies and their corresponding amplitudes. The output of each FFT constitutes the input features for our machine learning models. 
Our ML-based classification model uses the popular XGBoost classifier \cite{xgboost}. XGBoost is a gradient boosting approach that uses an ensemble machine learning model algorithm, and the ensembles are based on decision trees. XGBoost is well known for its computational efficiency and model performance. We train an XGBoost model using training data gathered from an academic building on our campus. Section \ref{sec:setup} describes the dataset used for training in more detail. 
To design \name's regression model, we first gathered ground truth airflow rate data using flow meter sensor attached to different vents and also captured audio samples from a microphone for each vent. While the amplitude of low-frequency audio signal increases with flow rate (e.g., higher airflow rate results in louder vent noise), we found that the relationship between airflow rate and the amplitude of the frequency components in the captured audio signal is non-linear. Hence, linear regression is not a suitable model choice. \names uses XGBoost regression instead, which can handle non-linear relationships between the input features and the output. We train an XGBoost regression model using our ground truth training data and deploy it for predictions. The resulting model uses the seven input features to predict the airflow rate from the vents. Next, we discuss how our ML model-based approach should be enhanced to be robust to interference and non-intrusive by avoiding human speech privacy leakage. 

\subsection{Enhancing Privacy and Accuracy through Silence Period Detection}
\label{sec:silentperiod}

Our ML approach presented above assumes that the audio sensing of the ambient environment is continuous. However, there are two drawbacks of continuous sensing. First, it will capture human speech in the surroundings, which will potentially leak privacy. Note that our low pass filter removes all the frequencies above 375 Hz, which largely removes spoken sounds. While speech information is largely concentrated in mid-frequency bands, some speech information is also present in low-frequency bands (see Figure \ref{fig:intuition_spectrum}). Hence, the audio signal will still contain some human speech even after the low pass filter stage.
Second, other ambient sounds in the environment including office equipment, noise from morning in the environment, outside traffic noise, nature sounds are also present in the audio signal. These ambient noises can also interfere with the sensing of airflow, as denoted in Figure \ref{fig:intuition_spectrum}.\\

\begin{figure}[ht!]
    \centering
    \includegraphics[width=0.6\textwidth]{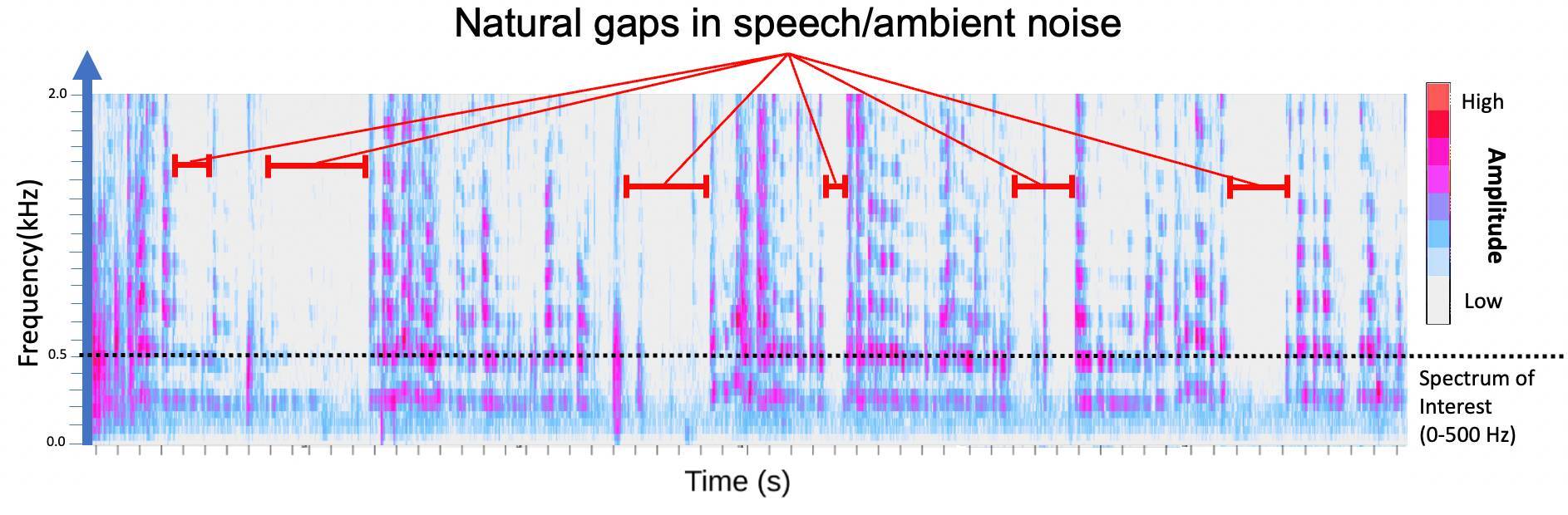}
    \caption{Audio frequency spectrum of three-way conversation and other ambient noise in an office room shows presence of silent gaps.}
    \label{fig:noisegap_spectrum}
\end{figure}

To address these twin issues of privacy and interference, \names employs a technique to detect silence periods, and performs sensing only during such silence periods.  Our silence period detection is effective for two reasons.
First, audio sensing of airflow through vents need not be continuous since airflow rate changes very slowly. Rather than sampling continuously, sampling every few seconds or minutes will yield the same information without any reduction in monitoring accuracy.
Second, neither human speech nor ambient sounds are continuous. There are gaps between spoken words and brief silence periods when humans are talking and interacting with one another. Similarly, there can be short or long gaps between sounds resulting from indoor or outdoor activities.  This can be seen in Figure \ref{fig:noisegap_spectrum}, which shows the audio signal captured from an office room with a three-way conversation and other typical background ambient noise.
As can \review{be} seen there are nature silence periods of varying duration in the audio signal. 


Hence, our approach focuses on detecting such short silence periods and capturing the audio signal only during such periods of relative silence. In doing so, it reduces the chance of capturing human speech components that are leftover after low-pass filtering, further enhancing privacy and review{reducing} interference from ambient noise. Note that our FFT transform works over 16 ms audio samples, so the silence periods can be relatively short (tens of milliseconds) for the approach to work well.
We use a threshold-based approach to detect a silence period.
To compute the silence threshold, we compute the maximum root mean square (RMS) value
of the noise generated by air vents (e.g., during our experimental data collection) and use that RMS threshold as a cut-off value. Audio signals that are "louder" than this cut-off are ignored and those below are deemed as silence periods and captured. Note that silence period detection is a pre-processing step and is performed {\em before} the low-pass filtering and ML prediction stages, 
as shown in Figure \ref{fig:sysOverview}.

A challenge in silence period detection is an appropriate choice of the silence threshold. The threshold  depends on the ambient environment and needs to \review{be} chosen carefully. If it is set too high, it can still capture faint sounds, including human speech occurring at a distance. If it is set too low, it can reduce the ability to capture audio signals with information about the airflow. Since the subsequent low-pass filtering stage  removes most frequencies corresponding to human speech, we make a design choice of using a higher threshold (e.g., by using the loudest vent observed in our data sets) to avoid losing any important information about the airflow and relying on the low pass filter stage to further remove any remaining human speech or other ambient noise in the captured silent audio signal. 
Our privacy experiments in Section \ref{sec:privacy} show that this is an effective trade-off that does not leak privacy.

\subsection{Ensuring Robustness through Minimum Persistent Sensing}
\label{sec:robustmps}


While silence period detection and sampling reduce interference from low-frequency noises, it does not entirely remove interference---other low amplitude (low volume) low-frequency background noise may still be present during silence periods. 

\begin{figure}[h!]
    \centering
    \includegraphics[width=0.5\columnwidth]{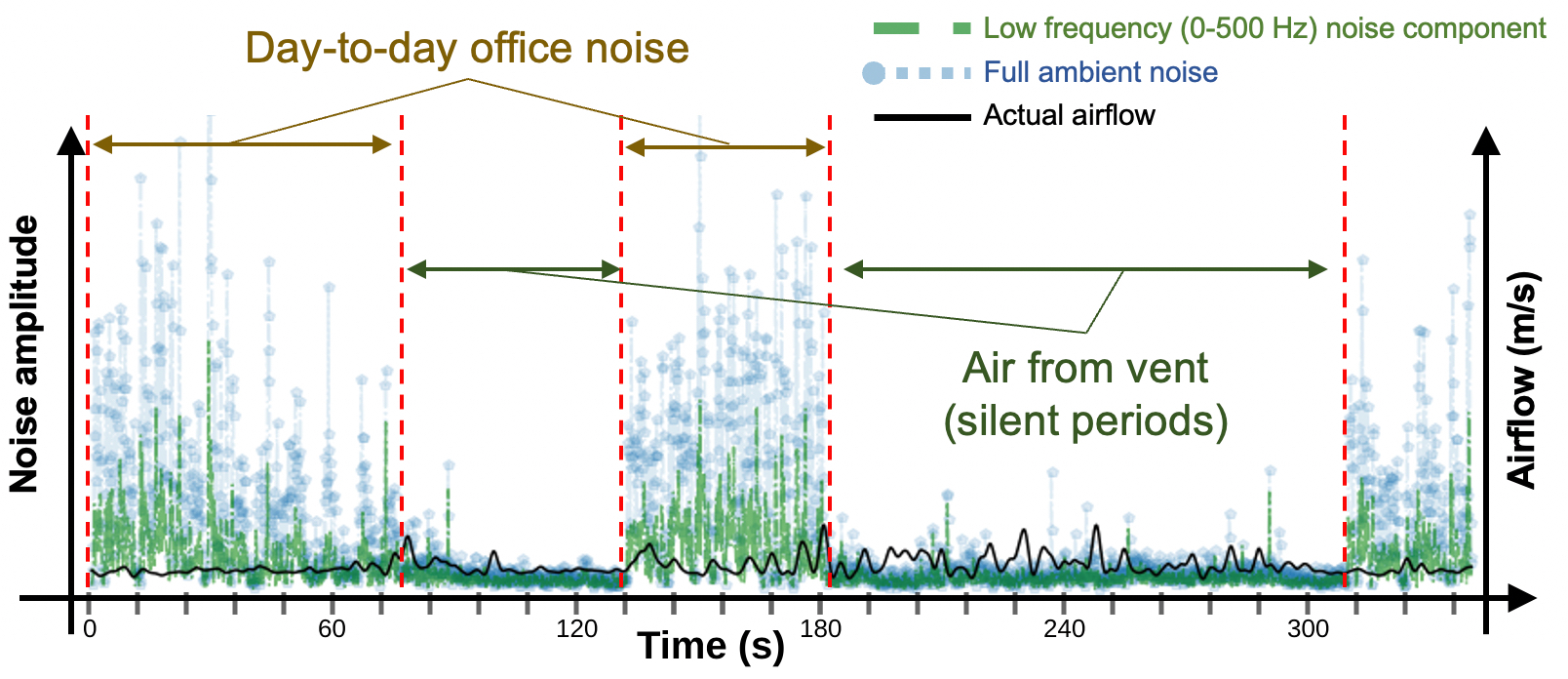}
    \vspace{-0.5em}
    \caption{Amplitude of the audio signal in an office room. The low-frequency audio, shown in green, remains steady when only vent noise is present and exhibits intermittent variations when other background noise is observed.   }
    \label{fig:intuition_ambientnoise}
\end{figure}

Our second enhancement, called Minimum Persistent  Sensing (MPS), is designed to improve the robustness of \names to low-frequency background noise. The main hypothesis behind MPS is that noise from airflow through the vents is relatively constant when the vent is on. In contrast, ambient background noise is intermittent and variable. Consequently, the amplitude of the low-frequency signal after low pass filtering will consist of two components---a relatively stable value from steady, continuous noise from the vents and a variable component ``overlaid'' on this signal due to intermittent and variable background noise. This behavior is visually depicted in Figure \ref{fig:intuition_ambientnoise} \review{shows the amplitude of the overall spectrum as the primary y-axis, the amplitude of low-frequency audio after low pass filtering as the secondary y-axis and airflow values over time for the x-axis collected using a smartphone microphone.} 
As can be seen, the amplitude of the low-frequency audio signal is steady when only vent noise is present and shows variability \emph{above} this minimum value due to intermittent background noise. 

If these intermittent variations are included during ML prediction, our models will over-predict the rate of airflow. Hence, our MPS involves determining a period where the signal exhibits a steady value with slight fluctuation and only considers ML predictions in such periods, preventing ambient noise from causing the model to over-predict the airflow rate. As shown in Figure \ref{fig:sysOverview}, MPS is a {\em post-processing} step that involves analyzing the predictions of our ML regression model to remove potential over-predictions and retain predictions indicating the minimum persistent value.
 
\begin{algorithm}[H]
\caption{Minimum Persistent Sensing (MPS) }\label{alg:euclid}
\begin{algorithmic}[1]
\State Initialize $n$, $\epsilon$, $p$

\Procedure{MinimumPersistingValue}{$arr$}\Comment{$arr$ is list of $n$ consecutive predictions}
\State $arr \gets sort(arr)$
\State $per \gets [arr[0]]$ \Comment{initialize the persistent sequence with the 1st element}
\State $i\gets 1$
\While{$i\not=n$}
\If{$abs(mean(per)-arr[i]) <= \epsilon$}
    \State $per.append(arr[i])$
\ElsIf{$abs(mean(per)-arr[i]) > \epsilon$}
    \State $per \gets [arr[i]]$ \Comment{if this element deviates by more than $\epsilon$, start a new sequence}
\EndIf
\If{$length(per) = p$}
    \State return $mean(per)$
\EndIf

\State $i\gets i+1$
\EndWhile
\State return \textbf{failure}
\State
\EndProcedure


\State $i\gets 0$, $arr[n]$
\While{$i\not=n$}
\State Get $p\_value$\Comment{$p\_value$ is the predicted value given by the model}
\State $arr.append(p\_value)$
\State $i\gets i+1$
\EndWhile

\State MPV = \textsc{MinimumPersistingValue}(arr)
\end{algorithmic}
\end{algorithm}

To do so, we use three parameters a) $n$: window size specifies how many predictions we should use to find the minimum persisting value, b) $\epsilon$: difference parameter specifies the maximum deviation of persistent predictions from the reported mean, and c) $p$: persistent parameter specifies the number of times similar values have to persist in being identified as a valid prediction. 

Given $n$ values, the algorithm finds a minimum value that persists at least $p$ times within a threshold of $\pm\epsilon$. We first wait for $n$ consecutive predictions and sort the values in non-descending order. We iteratively build a persistent sequence -- starting from the first value, we check if the next value in the sorted list lies within $\pm\epsilon$ of the mean of our current sequence. If yes, we update the length of the persistent sequence to include this element, and if \review{not}, we restart the search and initialize a new persistent sequence consisting of the current value.  We terminate when the length of the persistent sequence is $p$, and report the average of the persisting values.  If a persistent sequence cannot be found, we do not report a predicted value and once again wait for $n$ consecutive predictions to try again.  This MPS algorithm is described in Algorithm 1.

\subsubsection*{Effect of MPS Parameters}
\review{Several considerations must be made in selecting the parameters for MPS. First is the number of samples, $n$. While more samples will result in more reliable results with higher accuracy, a larger $n$ will also cost more time in collecting samples (e.g., n=100 will result in acquiring 1.6 seconds of 16 ms audio samples) and computational time of sorting. Second is the persistent parameter $p$ which should always be less than $n$. A higher value of $p$ will result in the MPS algorithm requiring the predicted airflow value to persist for an extended time, which may not depict a realistic environment where variability in ambient noise exists. In an ideal case, $\pm\epsilon$, the third variable, should be low, as it represents the tolerance of variations in the persisting value.
}

\section{FlowSense Prototype Implementation}

We implemented two prototypes of \name, one on a low-cost Arduino microcontroller-based fixed sensing platform and the other using smartphones as a mobile sensing platform. This section describes our software implementation, which is common to both the fixed and mobile sensing platform, and then discusses our hardware prototypes.

\subsection{\names Software Platform}

Figure \ref{fig:sysOverview} depicts the software architecture of \name. We implement \names as a set of Python and Java modules. The modules are based on the Python \textit{scikit-learn} \cite{scikit-learn} framework for our Machine Learning algorithms and the \textit{kissfft} library \cite{kissfft} for efficient Fast Fourier Transform (FFT) computation. Our software platform consists of three key modules: (i) Sensing and Filtering module, which is used for sensing of audio data, (ii) Transformation module, which transforms the data to the frequency domain using FFT, and (iii) Prediction module, which implements \name's machine learning classification and regression models. Our platform also implements two key enhancements for privacy and robustness: silence period sensing and minimum persistent sensing (MPS). As shown in Figure \ref{fig:sysOverview}, silence period detection and sensing are implemented in the Sensing and Filtering module as a pre-processing step, while MPS is implemented as a post-processing step that is applied to the output of the regression model in the prediction module. Next, we describe each component in more detail. 
\newline

\noindent \textbf{Sensing and Filtering Module} \names begins by capturing audio signals from a microphone at the sampling rate of 16 kHz. We chose this sampling frequency since it is the lowest native sampling frequency supported on modern smartphones and our fixed Arduino-based microphone sensors. On Android, we use the built-in \textit{AudioRecord} \cite{googleaudiorecord} API to stream and buffer microphone input from the device's main microphone. On Arduino, we use the onboard microphone sensor MP34DT05 PDM (Pulse-Density Modulation) that uses to represent an analog signal with a binary signal. Since the captured audio signal may consist of ambient human speech, we subject the audio signal to silence period detection, which partitions the signal into $n$ ms segments and iteratively discards each segment that contains enough noise to classify as a ``non-silent'' segment.  In order to differentiate between silent and non-silent audio, we compute the RMS (root mean square) value of the segment and compare it against a silence threshold. To find this threshold, we chose a value based on our data set as detailed in Section \ref{sec:silentperiod} -- this threshold is an RMS value of $silent\_threshold=60$. All segments with audio levels below our silence threshold are then subjected to low-pass filtering as shown in Figure \ref{fig:sysOverview}. The low-pass filter removes all medium and high components in the audio signal, including any residual (or faint) human voices present in silence periods. As discussed in Section \ref{sec:cut-off}, \names uses a cut-off frequency of 375 Hz for the low-pass filter, which removes any residual speech as well as other high-frequency noises. \newline

\noindent \textbf{Transformation Module} The resulting low-frequency audio signal is then transmitted using FFT that we compute using the \textit{Noise} wrapper, which implements \textit{kissfft} \cite{noise} natively on Android. For Arduino, we use ArduinoFFT \cite{ArduinoFFT} library to transform the signal to frequency domain.  Like many real-time digital signal processing applications, the FFT must be computed on a shorter duration. In our case, we apply FFT to 256 audio samples at a time. Since our sampling rate is 16 kHz, this means each FFT is applied to a 16 ms (256/16kHz) audio segment, which then yields a frequency domain signal containing the seven audio frequency ranges of interest (from 0-375 Hz) and their amplitudes.  These seven frequency ranges and their amplitudes serve as the input features for our models. 
\newline

\noindent \textbf{Prediction Module} The prediction module runs our machine learning models. These models are trained offline and then deployed in our software platform to predict vent state and airflow rate in real-time. The popular scikit-learn framework \cite{scikit-learn} is used for offline training as well as online predictions. The module executes two ML models concurrently. The first model predicts the state of the air vent using binary classification to predict a discrete label of on or off. As discussed in Section \ref{sec:MLmodels}, we use the open-source XGBoost library \cite{chen2016xgboost} for binary classification. The second model uses regression to translate the amplitude of the audio frequencies (``volume'' of noise through the vents) to predict the airflow rate. As discussed, since the relationship between amplitude and flow rate is non-linear, we use XGBoost for regression since it can handle non-linear relationships. The output of the regression model is subjected to the MPS to detect a stable minimum rate, which is then output as the rate of airflow. \review{Our parameter selection for MPS, as discussed in Section \ref{sec:robustmps}, led to us employing a random search technique evaluating a combination of parameter space to yield high accuracy and minimal time. The final values are as follows: $n=25$, $p=5$, and $\epsilon=0.5$.}


Overall, our implementation of silence detection and MPS enables \name’s ML models to produce useful predictions even in the presence of audible and frequent ambient noise.

\subsection{ \names Hardware Prototype}

The \names software platform is designed to run on two hardware prototypes of \names that we have built.

\begin{figure}[h!]
    \centering
    \includegraphics[width=\linewidth]{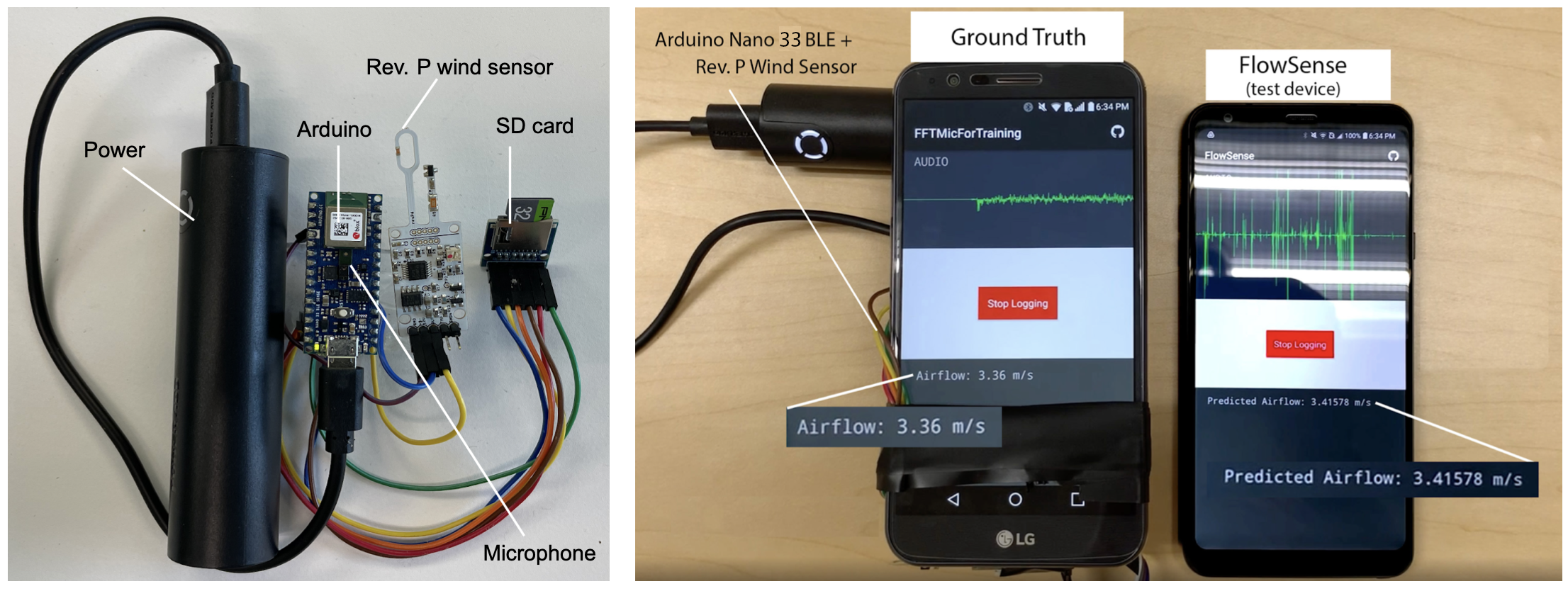}
    \vspace{-1em}
    \caption{\names Arduino and smartphone implementations}
    \label{fig:deviceExp}
\end{figure} 

Our first hardware prototype is designed for low-cost sensing using fixed infrastructure. \review{The cost of our fixed sensing prototype is approximately \$50 (i.e., Arduino with onboard microphone, \$35, SD card  and reader, \$8), while the mobile application is at no cost from leveraging smartphone capability. Contrasting the cost of our device fixtures with existing commercial airflow sensing devices (e.g., Honeywell AWM720P1 Airflow Sensor\cite{honeywellairflow} at \$200), our implementation is four times less expensive. However, it is important to note that our current implementation of \names does not include the additional capabilities supported by existing commercial devices. In Section \ref{sec:discussion}, we discuss this as a limitation.} As shown in Figure \ref{fig:deviceExp}, the prototype is based on Arduino Nano 33 BLE Sense platform \cite{arduino2019} with an onboard microphone. The Nano is powered using a generic external battery bank \cite{batteryBank} and logs all data on an SD card. We also use the same Arduino platform for ground truth data collection. When deployed for ground truth data collection, as opposed to real-time prediction, it is equipped with an additional Rev P wind sensor \cite{moderndevice2020} that we attach to air vents. The sensor can directly measure the rate of airflow through the vent, which we can then use a ground truth for the audio data collected using the microphone. In our current prototypes, several Arduino Nano devices can be deployed in different rooms or different parts of larger rooms. The devices can perform silence detection and FFT on the device. The FFT output can be logged to a file or transmitted over WiFi to a more powerful node such as a Raspberry Pi \cite{rpi} to perform the final ML prediction step.

\begin{figure}[!h]
\centering
    \includegraphics[width=.7\linewidth]{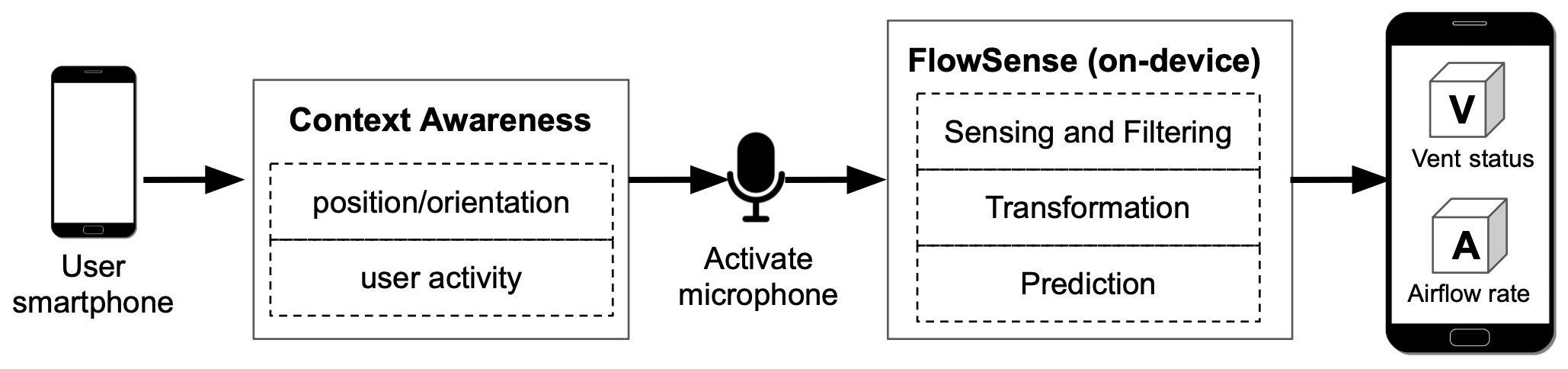}
    \caption{Smartphone implementation of \name.}
    \label{fig:mobileOverview}
\end{figure}

Our second prototype uses an Android \cite{android} smartphone as mobile sensor. We have implemented our entire \names software platform on an Android app. We use the \textit{JPMML} \cite{jpmml} project to convert our Python trained model into a Java-compatible object that we can run directly on Android. Figure \ref{fig:deviceExp} depicts the \names app running on an Android phone. As shown in Figure \ref{fig:mobileOverview}, our smartphone implementation also uses a context module that detects device movement, orientation, and phone exposure. The goal of the model is to activate the microphone only when the phone is stationary, placed in an orientation facing up (e.g., towards the ceiling vents), and exposed to ambient air (e.g., not in a bag or pocket).

As we show in Section \ref{sec:devicepos}, \names yields accurate results only when the microphone is unobstructed and faces the air vents. Our context-aware module infers the phone's exposure, movement, and orientation using the proximity, light sensor, and accelerometer APIs that are accessible through Android's \textit{SensorManager} framework \cite{googlesensormanager}. Once the context-aware module activates the microphone, the captured audio signal is subjected through the full \names software pipeline shown in Figure \ref{fig:sysOverview}.

Our prototype smartphone app runs on Android 6.0+ and has been deployed on five different smartphones; hardware specifications are detailed in Section \ref{sec:setup}. The phones that we deployed and tested the app on are a holistic cross-section of the Android installed base, spanning each OS version from 6.0 to 10 and representing a wide variety of hardware capabilities.
\section{\names Evaluation}
\label{sec:finalEval}
In \review{this} section, we evaluate the efficacy of \name's machine learning models and \review{their} robustness to various real-world interference. We also evaluate our privacy-preserving pipeline to sense airflow in a non-intrusive manner. 

\subsection{Experimental Setup}
\label{sec:setup}

We begin by describing our experimental setup and datasets, which are summarized in Table \ref{tab:studysummary}, and also describe our evaluation metrics.

\begin{table}[h]
\centering
\caption{\finalrev{Summary of Experimental Setup}}
\label{tab:studysummary}
\vspace{-1em}
\scalebox{.83}{
\begin{tabular}{|ll|ll|}
\hline
    \textbf{Dataset} & \makecell[tl]{Arduino: 80 x 30-min clips \\Smartphone: 5 phones x 6-hours clips \\ Human speech: 2 readers x 1-min clips} & \textbf{Deployment Duration} & Two weeks \\ \hline
    
    \makecell[tl]{\textbf{Environment}} & \makecell[tl]{Conference room (controlled) \\ Classroom, Laboratory \\ Bedroom} & \makecell[tl]{\textbf{Vent Types}} & \makecell[tl]{Square-like ceiling vent (controlled) \\ Linear sidewall vent, Linear ceiling vent} \\ \hline
    
    \makecell[tl]{\textbf{Fixed Hardware}} & \makecell[tl]{Arduino Nano 33 BLE Sense\\Rev. P wind sensor} & \makecell[tl]{\textbf{Mobile Hardware}} & \makecell[tl]{Tr-1: LG Stylo 4, \\Tt-1: Samsung Galaxy S8, \\Tt-2: Google Pixel XL, \\Tt-3: OnePlus One, \\Tt-4: LG Stylo 3 Plus} \\ \hline

\end{tabular}} 
\end{table}

\subsubsection{\finalrev{Data Ethics \& IRB Approval}}
Our data collection to experimentally validate the efficacy of \names has been approved by our Institutional Review Board (IRB). With our prototypes deployed in the wild, this process ran over two weeks under various indoor occupancy conditions. It is important to note that \names did not collect any audio signals with human speech. We stored only processed files where mid and high-frequencies signals were discarded, including suppressing human speech within our system's sensing and filtering module. As part of our privacy evaluation in Section \ref{sec:privacy}, we separately created audio clips with human speech. 

\subsubsection{\finalrev{Environment}} We deployed our \names prototypes in two office buildings and one residential building with central HVAC systems.  \review{Note that the building ventilation in our test environment is based on fixed schedules and not driven by human occupancy.} We use these deployments to evaluate \names in a controlled environment as well as real-world settings. For controlled experiments, we use a mid-size office with a capacity to house 6-8 occupants. The room has two ceiling air vents positioned at two opposite ends of the room controlled by the building's HVAC and building management system. The room allows careful control of the number of occupants and ambient noise. For our real-world experiments, we deployed \names in a different office building with a different HVAC system and the same type of ceiling air vents along with side vents. We also deployed \names in a residential building with a central HVAC system and different types of ceiling vents. 
These deployments simulate real-world conditions since we perform measurements with regular occupants of those spaces and typical ambient noises such as conversation, digital music, office machines, and outside city noises.

\subsubsection{\finalrev{Device}} As shown in Figure \ref{fig:deviceExp}, we deploy both our fixed and mobile sensor prototypes in the above spaces. For fixed sensing and ground truth data collection, we use an Arduino Nano 33 BLE Sense \cite{arduino2019} with an onboard microphone sensor and Android smartphones. These devices are \textbf{Tr-1} (LG Stylo 4, Snapdragon 450 @ 1.8 GHz, 2GB RAM, Android 8.1), \textbf{Tt-1} (Samsung Galaxy S8, Snapdragon 835 @ 2.35 GHz, 4GB RAM, Android 9.0), \textbf{Tt-2} (Google Pixel XL, Snapdragon 821 @ 2.15 GHz, 4GB RAM, Android 10), \textbf{Tt-3} (OnePlus One, Snapdragon 801 @ 2.5 GHz, 3GB RAM, Android 6.0) and \textbf{Tt-4} (LG Stylo 3 Plus, Snapdragon 435 @ 1.4 GHz, 2GB RAM, Android 7.0). We designed the app to run locally on-device to avoid transmitting audio data, thus guaranteeing the privacy of the proposed system. We utilized the Rev. P wind sensor by Modern Device \cite{moderndevice2020}, which is capable of detecting wind direction and speed. Using an SD card reader module and SD card to store the data files, the Rev. P wind sensor is solely for collecting ground truth.





\subsubsection{\finalrev{Datasets}}
We gathered data over two weeks in the above environments. The Rev. P wind sensor generates airflow rates in meters per second, which we use as labels for our regression model. We bucketize airflow rates into \emph{1-- on} when the air vent is running, otherwise \emph{0-- off}, for our classification model. Our Arduino training dataset contains data recorded at different locations in the controlled environment setting, amounting to 80 30-minute long clips recorded for over two weeks, each having roughly 60000 samples. This includes files recorded for different distances and orientations of fixed setup from the vents in the controlled environment setting. The testing dataset, however, included 30 minutes long audio clips from different vents.

Unlike the fixed implementation of Arduino, our smartphone dataset consists of low-frequency audio clips sensed for airflow using five different smartphones. Test clips are estimated to be 30 minutes long (roughly 15,000 samples each), under various smartphone placements and orientations from the air vent.  We also collect training audio clips using one smartphone (\textbf{Tr-1}) under various distances and orientations, where clips are approximately 6 hours long (roughly 300,000 samples each). These variations in our data are to account for smartphone users owning different device models and holding their devices in many positions. Altogether, our data set contains roughly 10 million samples (recorded over a week duration) -- $\approx5$ million samples are classified as \emph{1-- on}, and $\approx5$ million samples are classified as \emph{0-- off}. We utilize a fraction of this data set to train our models (80\% of data from \textbf{Tr-1}), and use the rest for evaluation.

\subsubsection{\finalrev{Evaluation Metric}}
Accuracy is the most intuitive measure for our classification model, with our dataset being somewhat balanced. We prioritize recall so that our model can accurately predict the air vent running when it is actually turned on. High precision relates to us predicting the air vent switched off when it is not running. F1-score is the weighted average of precision and recall. Next, we utilize the mean-squared error (MSE) for our regression model. MSE is the average of the square of the difference between actual and estimated airflow values. Additionally, we use the regression score, ${r}^2$, which is the coefficient of determination of the prediction. Ideally, MSE should be small and close to 0, and its range is dependent on the output value. Similarly, the most optimal ${r}^2$ is 1.0 and can be negative with poor model performance. ${r}^2$ is not always a good measure, particularly in cases when the test data have \review{less variance} (i.e., ${r}^2$  score is the ratio of the sum of squared errors to the variance in the dataset). As such, we prioritize MSE as our metric of performance for the regression models.

\subsection{Efficacy of ML Models}
\label{sec:MLmodels}
Our first experiment compares \names employing different classifier and regression approaches in a controlled experimental setting. Specifically, we compare \names to several standard algorithms, including Logistic Regression (classification), Linear Regression (regression), K-Nearest Neighbor (k=5), Support Vector Machine (SVM), Decision Tree (DT) with maximum depth = 5, and Gradient Boosting (XGBoost). As shown in Table \ref{tab:mlperformance}, XGBoost yields comparable accuracy in classifying vent status at 99\% to other ML algorithms (differences are not significant). In contrast, the performance improvement in predicting the rate of airflow is significantly better at ${r}^2$=0.96 using XGBoost than Linear Regression at ${r}^2$=0.49 (p<0.001). While the differences between XGBoost and SVM/DT are not significant (p>0.05), our decision to employ XGBoost is also because of its added advantages of low complexity, where it can easily run on low-end edge devices in real-time \cite{chen2015xgboost}.
\begin{table} [!h]
\caption{\finalrev{Classification and Regression performance on different  ML algorithms.}}
\label{tab:mlperformance}
\vspace{-1em}
\scalebox{.9}{
\centering
  \begin{tabular}{|l|c|c|c|c|c|c|c|c|}
    \hline
    \multirow{2}{*}{\makecell{\textbf{ML Algorithm}}} &
      \multicolumn{4}{c|}{\textbf{Classification}} &
      \multicolumn{4}{c|}{\textbf{Regression}} \\
      & Acc & Prec. & Rec. & F1 & Train MSE & Test MSE & Train Reg. & Test Reg. \\ \hline
      Logistic Regression & 0.98 & 0.98 & 0.98 & 0.98 & - & - & - & -\\
      \hline
      Linear Regression & - & - & - & - & 80.05 & 70.57 & 0.57 & 0.49\\\hline
      Support Vector Machine & 0.99 & 0.98 & 0.99 & 0.98 & 33.23 & 13.57 & 0.77 & 0.74 \\\Xhline{2\arrayrulewidth}
      k-Nearest Neighbor & 0.99 & 0.98 & 0.99 & 0.99 & 6.81 & 1.85 & 0.90 & 0.90 \\\hline
      Decision Tree & 0.99 & 0.97 & 0.99 & 0.99 & 0.51 & 1.91 & 0.93 & 0.91 \\\Xhline{2\arrayrulewidth}
      \rowcolor{LightCyan}
      XGBoost & 0.99 & 0.98 & 0.99 & 0.99 & 0.31 & 1.57 & 0.96 & 0.92 \\\Xhline{2\arrayrulewidth}
  \end{tabular}}
\end{table}

\subsubsection{Selecting the Cut-off Frequency}
\label{sec:cut-off}
Our implementation must select the most optimal cutoff frequency for our low-pass filter since the cutoff frequency is a trade-off between achieving high accuracy and maintaining user privacy. Our empirical observation in Section \ref{sec:motivation} observes the audio signal spectrum of air sounds from the HVAC ranging between 0-500Hz, thus, informing the cutoff frequency threshold for our low-pass filter. 

\begin{figure}[h!]
    \centering
    \includegraphics[width=0.6\linewidth]{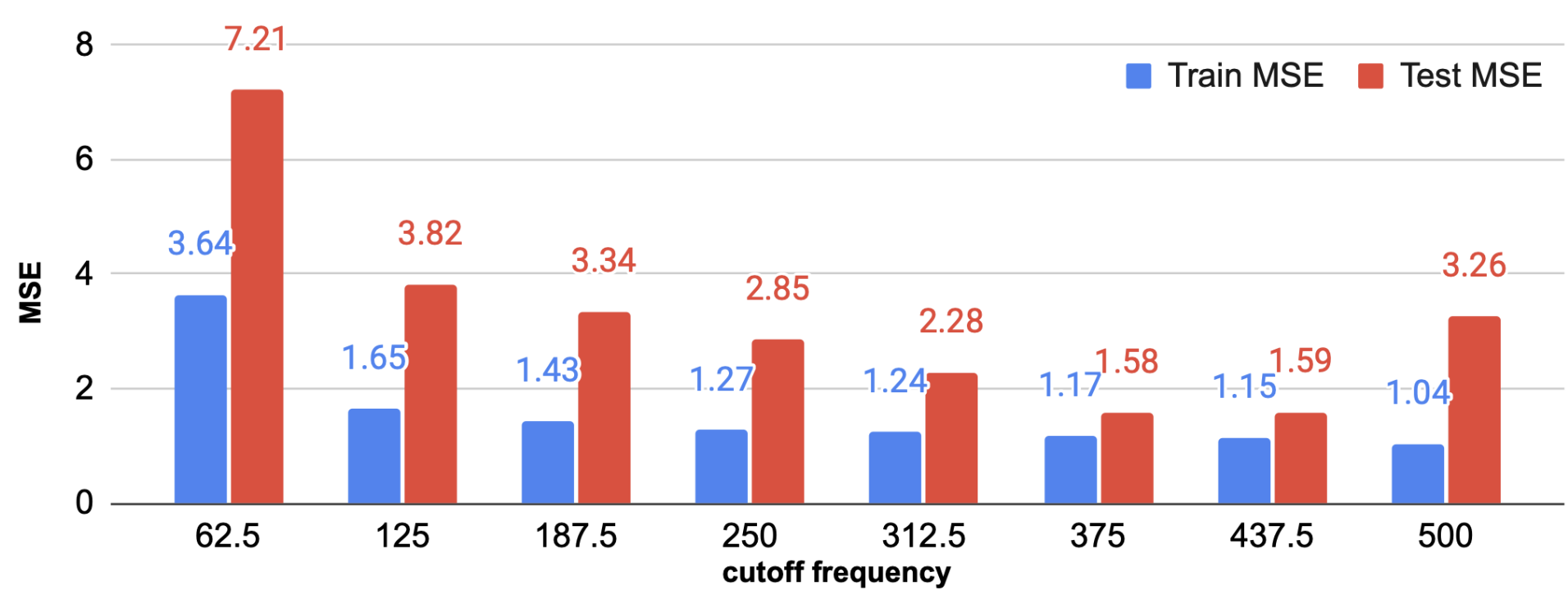}
    \caption{Determining ideal cutoff Frequency for the low-pass filter.}
    \label{fig:cutoffFreq}
\end{figure}

As shown in Figure \ref{fig:cutoffFreq}, training an XGBoost regression model with input data based on different cutoff frequencies can significantly impact model performance in predicting airflow rate. For example, while a cutoff frequency of 62.5Hz is most favorable to preserve user privacy (i.e., eliminate any audio signals above this range), both train and test errors are significantly higher by 3.64 and 7.21 MSE than preserving signals at 500Hz (p<0.001).  Our results found that a cutoff frequency of less than 312.5 will lead to above 2.0 MSE due to insufficient features during model training. On the other hand, preserving audio signals at 500Hz may result in system processing more ambient noise including, residual fragments of human voices. As a result, the dominance of this noise at 500Hz can lead to incorrect predictions with 3.26 MSE. Balancing accuracy and user privacy, we empirically decide on \emph{375Hz as the lowest threshold}, where the difference in error is lower than 5.63 MSE compared to 62.5Hz (p<0.001) and lower than 1.68 MSE compared to 500Hz (p<0.05).

\subsubsection{Model Generalization on Vent Types}
We evaluate \names in a natural setting to determine if our technique can be generalized to predict airflow from vent sources, located in different rooms and building types. These settings include conference rooms and a laboratory from two office buildings and a bedroom from a residential building. We test our model to continuously predict 30 minutes of airflow in each environment and summarize our findings in Table \ref{tab:vents}.

Our results \review{demonstrate} generalizability, particularly among buildings and rooms within the institution. However, the dimensions of the vent and its outlet geometry can significantly affect model performance. Specifically, we found that model performance on similar-typed vents -- square-like geometry outlet and on-ceiling -- is not significantly different from our controlled setting. Since the vents in the classroom of Building 1 is a sidewall vent with a linear-shaped geometry outlet, the model resulted in errors significantly more by 20.99 MSE (p<0.001). The geometry of the vent outlet in our residential setting is also different, resulting in 5.29 MSE, which is significantly higher than our controlled setting by 1.70 MSE (p<0.01).

\begin{minipage}{0.55\linewidth}
    \centering
    \vspace{2.5em}
    \captionof{table}{\finalrev{Performance of regression models for different vents}}
    \label{tab:vents}
    \scalebox{.85}{
	\begin{tabular}{|l|c|c|c|c|}
    \hline
        \textbf{Room Type} & \textbf{Building} & \textbf{Vent Outlet} &  \textbf{Occupancy} & \textbf{MSE} \\ \Xhline{2\arrayrulewidth}
       *controlled & Building 1 & square-like ceiling & 6-8 & 1.70 \\\Xhline{2\arrayrulewidth}
       Mid-size laboratory & Building 1 & square-like ceiling & < 20 & 1.92 \\
       Conference room & Building 2 & square-like ceiling & 6-8 & 1.84 \\
       Classroom & Building 1 & linear sidewall & < 50 & 22.69 \\
       Bedroom & Residential & linear ceiling & 1-2 & 5.29 \\ \hline
    \end{tabular}}
    \vspace{1.5em}
\end{minipage}\hfill
\begin{minipage}{0.45\linewidth}
	\centering
	\includegraphics[width=.65\linewidth]{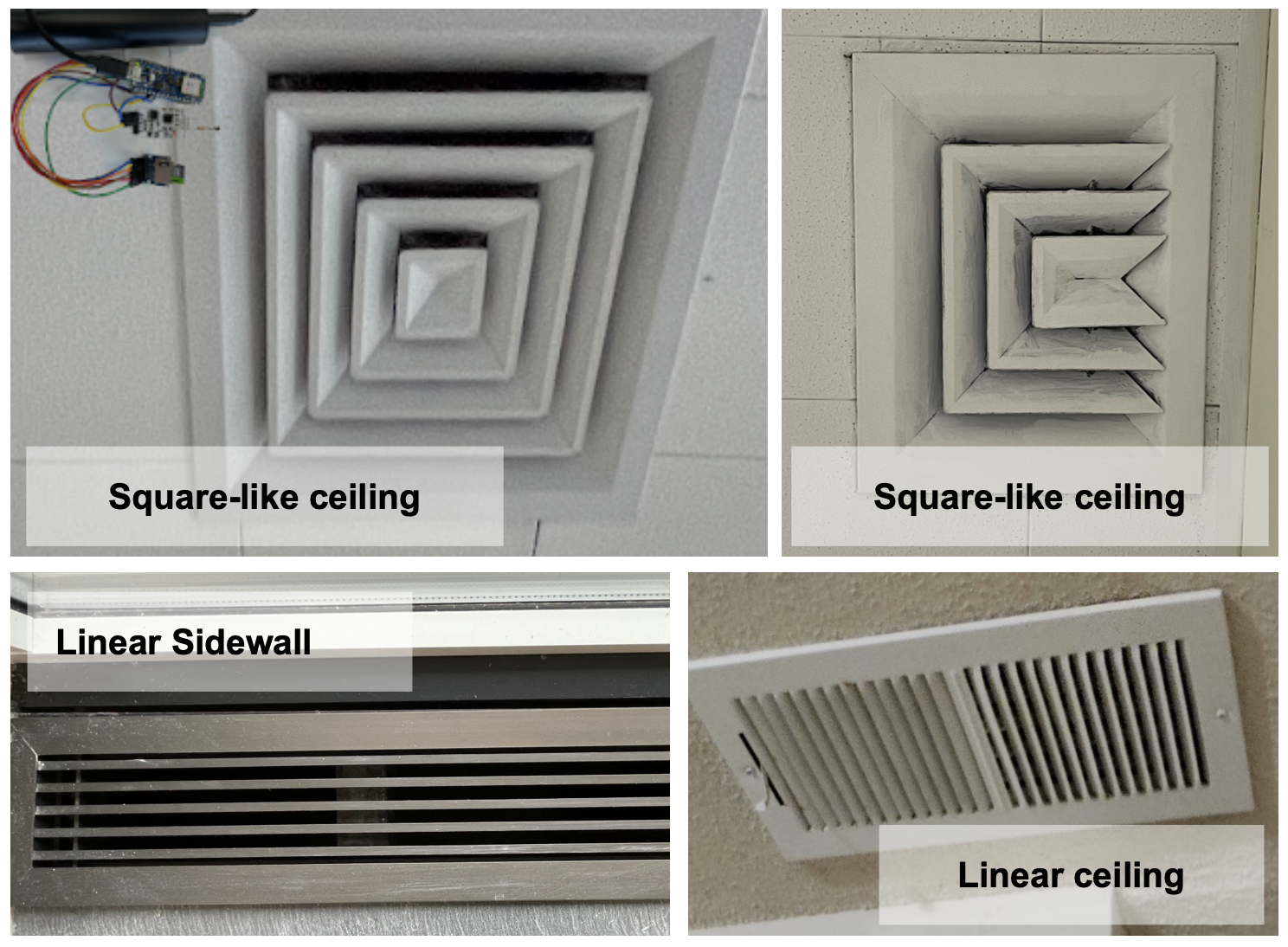}
	\vspace{-1em}
    \captionof{figure}{Types of vent outlet}
    \label{fig:ventoutlet}
\end{minipage}



\noindent\textbf{Key Takeaways}
Our experiments demonstarte the efficacy of \names ML models. Specifically, Gradient Boosting (XGBoost), which outperforms other standard ML models in model complexity, achieves 99\% accuracy in predicting vent status and ${r}^2$=0.92 in predicting airflow rate. Further, our experiments found the ideal cutoff frequency for our low-pass filter to 375Hz, discarding audio signals above this frequency. In doing so, we can preserve enough frequency dimensionality to learn the patterns of audio spectrum from HVAC to predict airflow rate at low errors of 1.17 MSE. We tested \names in different locations, rooms, and vent types, concluding that while our model is generalizable to room dimensions, it is significantly impacted by the geometry and placement of the vent outlet. 

\subsection{Robustness to Interference}
Next, we evaluate the robustness of \names in handling real-world  challenges arising from interference caused by ambient noise present in the environment, positioning of the experimental setup, and user privacy.
The broader goal of our work aims to provide a system suitable for everyday use among everyday users. Everyday use, however, comes with several real-world interference challenges primarily arising from atmospheric noise and smartphone variations. The following experiments aim to understand better how \names performs under these conditions.

\subsubsection{Effect of Ambient Noise}
To evaluate robustness, we subject \names to different types of ambient noise and different noise levels. We consider several everyday indoor settings where \names will be utilized.  They are: (1) a shared environment with multiple people conversing (e.g., occupants gathered for a meeting), (2) an office environment with high-performing workers (i.e., office machines, laptop, and computer fans), (3) a personal environment with digital audio sources (e.g., music playing from the speaker, teleconferencing), (4) a personal environment with distant city and traffic sounds, and (5) a personal environment where user is walking.
\begin{table}[h!]
\caption{\finalrev{\name's performance in predicting airflow rate under different atmospheric noises.}}
\label{fig:ambientNoise}
\vspace{-1em}
\scalebox{0.9}{
\begin{tabular}{|c|l|l|c|c|c|}\hline
&\textbf{Environment} & \textbf{Noise Type/Source} & \textbf{dB level} & \makecell{\textbf{Naive Pred.}\\MSE} & \makecell{\textbf{Silence Period + MPS}\\ MSE}\\\Xhline{2\arrayrulewidth}
controlled & \makecell[l]{shared room} & ambient noise & 36 dB & 1.70 & 0.50\\\Xhline{2\arrayrulewidth}
env 1 & shared room & 2-3 people conversing & 56 dB & 21.91 & 0.65 \\
env 2 & shared room & white noise of office machines & 46 dB & 3.56 & 0.77 \\ 
env 3 & personal space & digital source of music and speech & 54 dB & 6.35 & 0.73 \\ 
env 4 & personal space & white noise of city noises & 50 dB & 3.30 & 0.44 \\
env 5 & personal space & white noise/user walking & 44 dB & 6.54 & 4.79 \\
\hline 
\end{tabular}} 
\end{table}

Recall in Sections \ref{sec:silentperiod} and \ref{sec:robustmps}, we proposed implementing the silence period detection and Minimum Persistence Sensing (MPS) algorithm to overcome challenges of inaccuracies resulting from ambient noise. Table \ref{fig:ambientNoise} summarizes our model performance in predicting the rate of airflow under environments exposed to everyday atmospheric noise, including human speech, digital sounds, and white noises, before (i.e., naive prediction) and after employing our proposed techniques (i.e., silence period + MPS). As discussed in section \ref{sec:silentperiod}, the presence of ambient noise negatively affects the naive predictions of the \names regression model by over-predicting airflow, including in our controlled environment. Overall, our results yield significantly lower MSE when the model employs silence period detection and MPS. For example, errors in our controlled setting reduced 1.20 MSE (p<0.05) with silence period detection and MPS. In a real environment of a small group conversing \review{(env 1), our technique effectively reduces error to 0.65 MSE compared to a naive prediction of 21.91 MSE (p<0.001), when speech is present. Note that the MSE of naive prediction is high in this case because the noise of people conversing resulted in intermittently high amplitude and decibel levels in the low-frequency spectrum. This problem is resolved by silence period detection and MPS. First, silence period detection does not allow speech above the silent threshold to pass through the system. Further, the effect of low-frequency speech having overall amplitude less than the silent threshold is removed by MPS. This function, however, will not perform in cases where continuous ambient noise above the silent threshold is present (e.g., crowded dining area, \names cannot detect a silent period to sample for audio signals). We discuss this shortfall in Section \ref{sec:discussion}.}


\begin{figure}[h!]
    \centering
    \includegraphics[width=.85\linewidth, height=4.4cm]{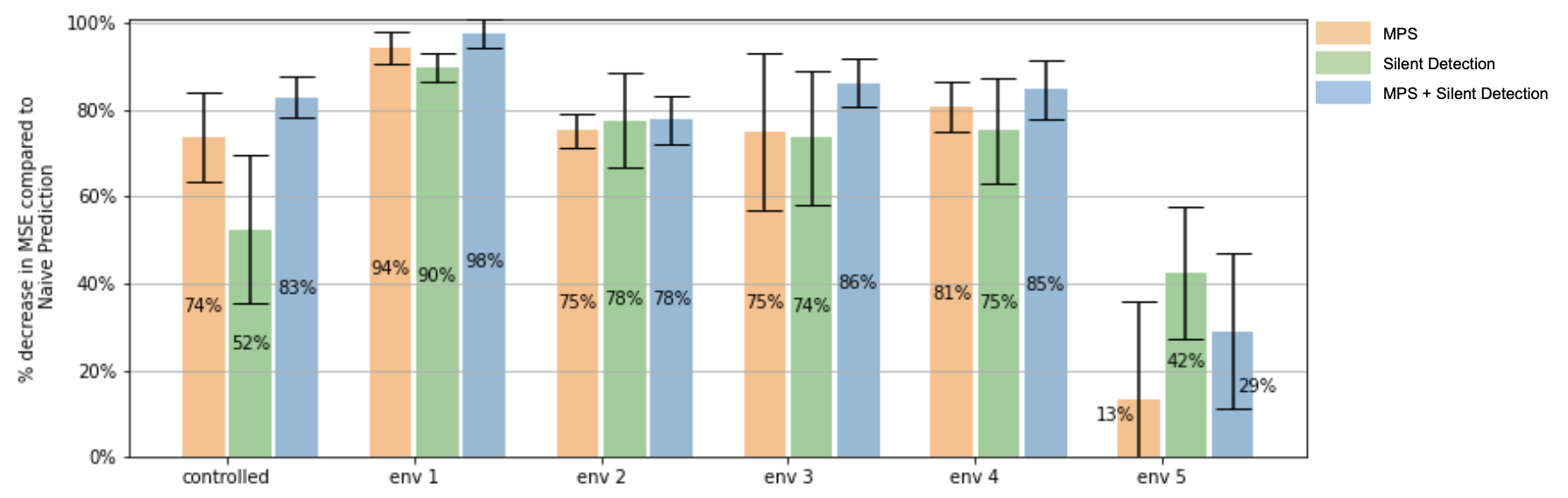}
    \caption{Performance of Regression Model with different cases of simulated and real ambient noise}
    \label{fig:casesambbar}
\end{figure}

\begin{minipage}{0.45\linewidth}
    \centering
    \vspace{1.5em}
	\includegraphics[width=\linewidth]{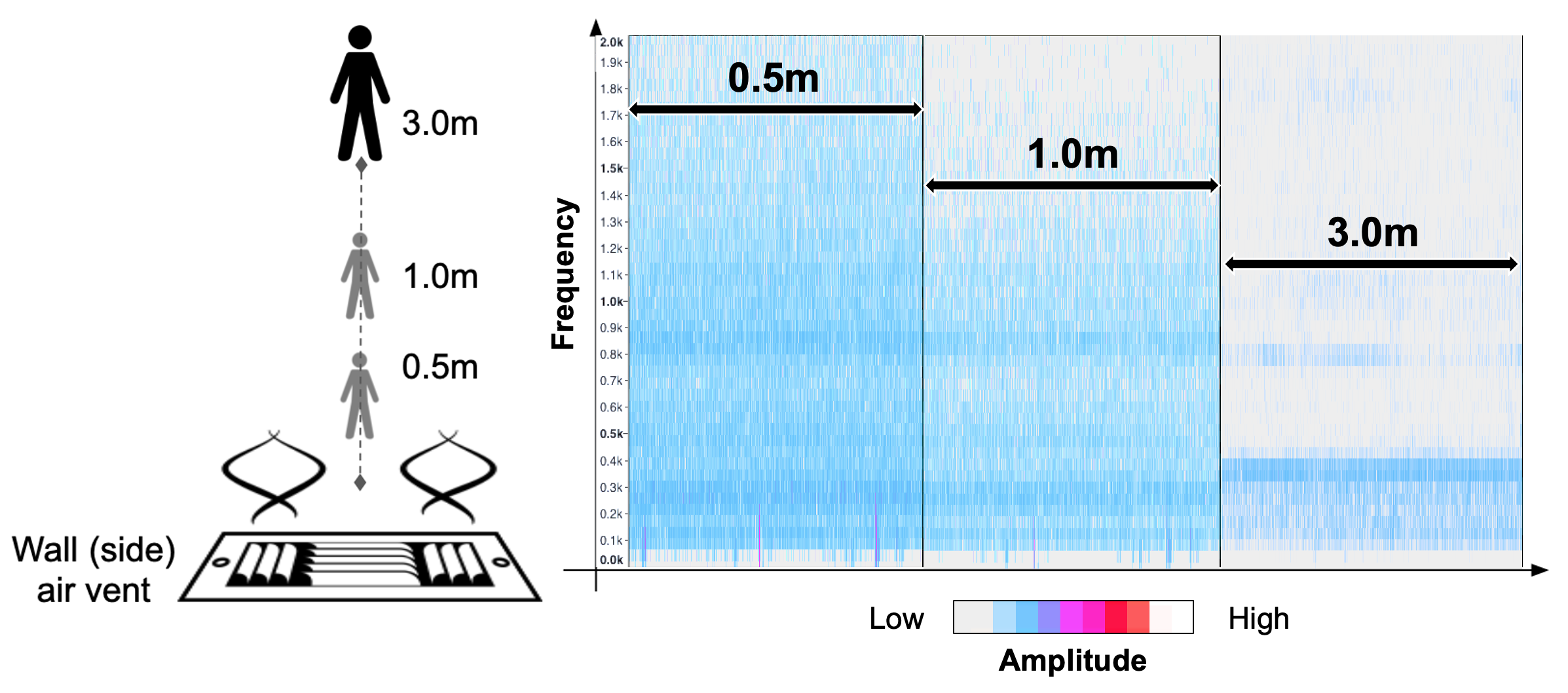}
    \captionof{figure}{The frequencies involved in the audio spectrum of airflow reduces with distance from its vent source.}
    \label{fig:intuition_amplitude}
\end{minipage}\hfill
\begin{minipage}{0.5\linewidth}
	\centering
	\includegraphics[width=.74\linewidth]{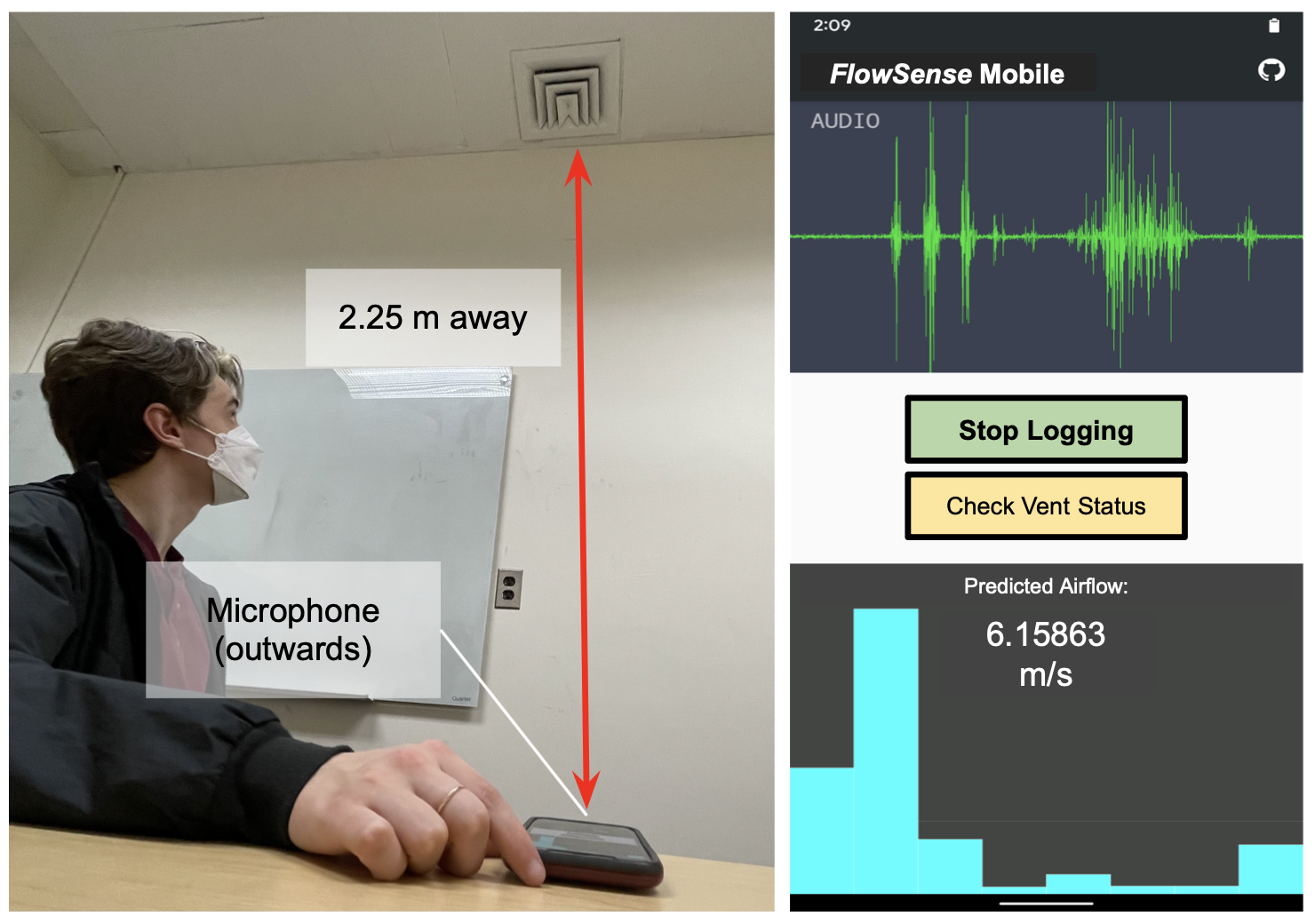}
	\vspace{-1em}
	\captionsetup{justification=centering}
    \captionof{figure}{Optimal device placement and orientation for user and screenshot of \name.}
    \label{fig:mobiledistance}
\end{minipage}\newline \newline

\noindent\textbf{Significance of Minimum Persistence Sensing} To better understand why we achieved these improvements, let us consider the results plotted in Figure \ref{fig:casesambbar}. Specifically, the figure shows the percentage improvements of our model predicting airflow utilizing either one of the techniques or collectively utilizing both techniques with naive prediction. Overall, our proposed enhancements of silence detection and MPS \review{improve} MSE by 77\% compared to naive prediction. MPS is most effective in reducing error under environments with prominent white noise, such as our controlled setting by 74\% and traffic-city noises by 81\% (env 4). It is important to note that our silence-detection algorithm mainly preserves user privacy rather than improves ambient noise. An anomaly in our results is from testing \names when the user is walking around the room (env 5) \review{with a smartphone in hand. Despite applying MPS, it reduces error by only 13\%. This result arose from large fluctuations in both ground truth airflow values and recorded audio. We observed that the speed at which the user is walking and grip of the phone could produce low-frequency noise from the phone moving. The inability to separate these low-frequency noises contributed to more significant errors, decreasing the effectiveness of MPS. The unpredictability of white noises from phone movement informs our decision to implement context-awareness as part of \names context-awareness as shown in Figure \ref{fig:mobileOverview}.} 

\subsubsection{Device Distance and Orientation }
\label{sec:devicepos}
It is most practical for users to hold their smartphones in any preferred way when utilizing \names app. Our next experiment investigates how different device \review{distances and orientations} affect our model performance. Distance is two points between the smartphone's microphone to the HVAC vent, with the nearest being 0.5 meters away. Orientation is relative to the frame of reference, in this case, the HVAC vent. 0° is the smartphone microphone facing towards the vent, 90° is the smartphone microphone facing along an axis perpendicular to the vent (e.g. given a ceiling-mount vent, the primary microphone faces one of the room's walls), 180° is the smartphone microphone facing away from the vent.
\newline

\noindent\textbf{Increasing Error with Distance} As shown in Figure \ref{fig:intuition_amplitude} above, since sound intensity is proportional to the square of the amplitude of waves, we can expect the amplitude of sound waves to decrease with increasing distance from the air vent source \cite{chowning1971simulation}. However, the maximum spatial boundary of our sensing technique is limited to before accuracy is compromised remains unknown.

\begin{table} [!h]
\caption{\finalrev{Performance of predicting vent status and rate of airflow at varying distances.}}
\label{tab:distorient}
\vspace{-1em}
\scalebox{.8}{
\centering
  \begin{tabular}{|l|c|c|c|c|c|c|c|c|c|c|}
    \hline
    \multirow{2}{*}{\makecell{\textbf{Orientation/Distance}}} &
      \multicolumn{5}{c|}{\makecell{\textbf{Vent Status}\\(Classification, Accuracy)}} &
      \multicolumn{5}{c|}{\makecell{\textbf{Rate of Airflow}\\(Regression, MSE)}} \\
      & 0.5m & 1.0m & 1.5m & 2.25m & 3.0m & 0.5m & 1.0m & 1.5m & 2.25m & 3.0m \\\Xhline{2\arrayrulewidth}
      0° (microphone upwards facing vent) & 0.983 & 0.957 & 0.945 & \cellcolor{LightCyan}0.942 & 0.821 & 0.63 & 0.72 & 0.73 & \cellcolor{LightCyan} 0.96 & 1.12
      \\ \hline
      90° (microphone perpendicular to the vent) & 0.960 & 0.944 & 0.889 & 0.953 & 0.942 & 0.62 & 0.84 & 0.73 & \cellcolor{LightCyan} 0.64 & 0.74\\ \hline
      180° (microphone downwards facing away from the vent) & 0.962 & 0.955 & 0.822 & 0.842 & 0.524 & 0.55 & 0.71 & 0.46 & \cellcolor{LightCyan} 0.57 & 0.95 \\ \hline
  \end{tabular}}
\end{table}

Table \ref{tab:distorient} summarizes our model performances in predicting vent status and airflow rate at different distances and orientations using \textbf{Tr-1}. Indeed, we observe that the amplitude of audio signal caused by air from the HVAC air vent to diminish with increasing distance -- in bucketizing vent status as on or off, it is more likely the decreasing amplitude of sound is \review{labeled} as `off' even when the vent is running. Our classification model is beyond 90\% accurate in informing users if the vent is running as long as the smartphone is within 2.25m (p<0.05) from the air vent—the increasing distance results in incremental drops. Classification accuracy drops at 82\% when the smartphone is placed 3.0m away from the vent. While results from our regression model \review{yield} slight inconsistencies between 0.63-0.97 MSE with increasing distance (at 0° orientation), the performance of our regression model is significantly impacted when the smartphone is placed 3.0m apart compared to when it is placed at 0.5m (p<0.01). That is, the regression error increases to 1.12 MSE. In a typical situation where users \review{place their smartphones on the table} roughly 2.25m away and facing the vent (0° device orientation), \names yields 0.96 MSE, in which the difference is non-significant compared to being 0.5m apart (p>0.05).
\newline

\noindent\textbf{Unobstructed Microphone Oriented Towards Vent}
Device orientation is also a factor that will affect our model performance, specifically for our classification model. When the smartphone is 2.25m away from the vent, we observe a significant drop of 10\% accuracy (p<0.05) by changing the microphone orientation from 0° to 180°. In contrast, our regression model is unaffected by device orientations because the model attempts to predict the airflow based on the audio signals \textit{in the immediate vicinity of the smartphone}. The prediction result is highly likely an under-prediction of the environment, however, it will be accurate to that of an obstructed airflow sensor. At 2.25m regardless of orientation, \names yields between 0.57-0.96 average MSE (differences are non-significant) to its performance at 0.5m.
\newline

These findings collectively imply that \names cannot operate accurately in predicting vent status and airflow rate when distance exceeds 3 meters or if the microphone is obstructed. In achieving high performance and, at the same time, balancing user convenience when utilizing \name, the phone or sensor should be 2.25m away from the vent, with the microphone sensor oriented facing the vent as shown in Figure \ref{fig:mobiledistance}. This measure is equivalent to a user standing below a ceiling vent with a phone in their hand for a 9 feet ceiling. The limitation of accurately sensing airflow when the microphone lies obstructed informs our decision to implement exposure detection as part of our context-aware capabilities -- see Figure \ref{fig:sysOverview}.

\subsubsection{Smartphone Variations} We expect different smartphone devices to affect our model performance since smartphones are likely to integrate different microphone specifications. For example, smartphones with microphones that are intended for voice-only recordings tend to have lower cutoff frequency, thus producing different \textit{frequency responses}, which may intuitively produce higher errors. Our experiment considers several smartphone models (i.e., \textbf{Tt-1}, \textbf{Tt-2}, \textbf{Tt-3}, and \textbf{Tt-4}), explicitly for test (of our trained model using \textbf{Tr-1}). 
\begin{figure}[h!]
  \centering
  \includegraphics[width=.8\linewidth]{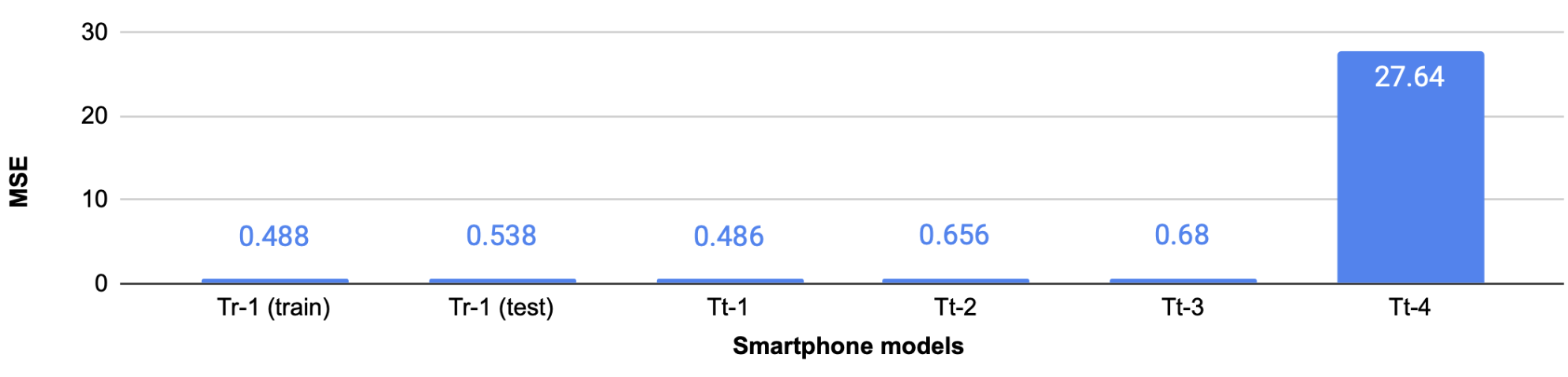}
  \caption{\name's model performance on different smartphone models.}
  \label{tab:summ_smartphone}
\end{figure}

Figure \ref{tab:summ_smartphone} tabulates \name's model performance on training and testing set using five smartphones. Overall, our results yield an average of  0.57 MSE in predicting the rate of airflow -- excluding \textbf{Tt-4} as the anomaly. The errors produced by other test smartphones did not exceed 0.2 MSE compared to errors produced by our training phone. In determining the cause of high error in \textbf{Tt-4}, we found that audio signals recorded by the phone were of moderately different noise amplitudes even though all phones were set in identical environments. Indeed, incompatibility arising from the difference in frequency response amongst smartphone models and their corresponding MEMS microphone parts is a known issue \cite{asmar2019, kardous2016, hermawanto2020}. While retraining models that match different frequency responses are required, these models can be generalized to different clusters of devices with similar microphone specifications. \newline

\noindent\textbf{Key Takeaway} In real-world settings, \names will be exposed to real-world interference phenomena, constituting atmospheric sounding and noise pollution to problematic usage of smartphones and model variations. Our experiments showed that \names \review{is} significantly robust to ambient noise upto 60 dB when utilizing our proposed techniques of silence period detection and MPS that improve the accuracy by 77\%. With fluctuating readings resulting from user movement and orientation, \names is enabled by context-aware capabilities to predict airflow optimally. However, users must be within 2.25m in the distance from the air vent. While \names is generalizable across most smartphone models, the make of the microphone sensor is significant in producing a similar frequency response at which our model is trained.

\subsection{Privacy \finalrev{Considerations}}
\label{sec:privacy}
As indoor environments are typically shared spaces, analyzing audio signals must prioritize user privacy. While our goal has been to develop \names as a privacy-preserving acoustic-based airflow sensor, it is not yet clear how the pipeline we have established is effective to actual users.

\begin{table} [!h]
\caption{\finalrev{Summary results of privacy evaluation questionnaire.}}
\label{tab:userprivacy}
\vspace{-1em}
\scalebox{0.8}{
\centering
  \begin{tabular}{|l|c|c|c|c|}
    \hline
    \multirow{2}{*}{\makecell{\textbf{Questionnaire}}} &
      \multicolumn{2}{c|}{\textbf{File A}} &
      \multicolumn{2}{c|}{\textbf{File B}} \\
      & Original & Filtered & Original & Filtered \\ \hline
      Were you able to hear anything in the file? & M=5.00,SD=0.00 & M=1.83,SD=1.52 & M=5.00,SD=0.00 & M=2.17,SD=1.47 \\ \hline
      
      \makecell[l]{Were you able to hear human sounds in the file? \\(i.e., sounds that could be coming from a human)} &M=5.00,SD=0.00  &  M=1.67,SD=0.39& M=5.00,SD=0.00 & M=2.08,SD=0.99 \\\hline
      
      \makecell[l]{Were you able to hear any speech in the file?} & M=4.83,SD=0.58 & M=1.16,SD=0.39& M=5.00,SD=0.00 & M=1.75,SD=1.05\\\hline
      
      \makecell[l]{Were you able to discern speech well enough to transcribe?} & M=4.75,SD=0.45 & M=1.00,SD=0.00 & M=4.92,SD=0.29 & M=1.08,SD=0.29\\\hline
      
      \makecell[l]{General comments: What did you hear if at all,\\what you perceived was in the file, can you transcribe?} & \makecell{talking about\\water flow,\\  movement of \\water masses} & \makecell{random noises,\\garbled sounds,\\white noise,\\mostly noisy} & \makecell{discussion of\\airflow, air press-\\ure, molecules} & \makecell{noise again,\\white noise\\\\}\\\hline
      
      What is your age? & \multicolumn{4}{l|}{M=25.09,SD= 4.72} \\\hline
      
      What gender do you identify as? & \multicolumn{4}{l|}{M:6, F:5, NB: 1} \\\hline

      What is your level of education? & \multicolumn{4}{l|}{7 Graduate} \\\hline
      
  \end{tabular}}
\end{table}

\subsubsection{User Study}
We conducted an IRB-approved user study to understand how effective our system is in preserving user privacy; this is evaluating the pipeline of sampling at 16kHz, using a low-pass filter (at 375 Hz cutoff frequency), and retaining audio signals upon successful silence detection.

Our study surveyed 12 participants (mean age=25) in a within-subject design to rate the clarity of hearing human speech and voices in two different audio clips (i.e., they are 30-second long readings by female and male readers). Note that these audio clips were recorded for the purpose of this user study. In our actual implementation, \names does not record and store audio. It only utilizes pre-processed low-frequency audio clips as input data. 

Table \ref{tab:userprivacy} tabulates the responses by our participants on a  Likert Scale questionnaire, rating 1: Not at all clear, 2: Somewhat unclear, 3: Neither clear nor unclear, 4: Somewhat clear, 5: Very clear. Overall, these participants confirmed the clarity of our input data as not sufficient to distinguish human speech. Participants rated an average mean score of 1.42 and 1.77 (somewhat clear) for the filtered Files A and B, respectively. Otherwise, original Files A and B were rated 4.9 (very clear), respectively.

\subsubsection{Privacy Evaluation \finalrev{Using} AI Speech Recognition Service}
Our evaluation also employed Google Cloud Speech Recognition as an NLP service \cite{googlenlp}. We use the \textit{SpeechRecognition} package for Python \cite{zhang2017speechrecognition} to detect speech from Files A and B (original and filtered) mentioned above. While the speech recognizer detected speech from the original files with a 95\% confidence score (A=0.965, B=0.95), the NLP service cannot detect any speech or possible translation from the filtered file. As a result of untranscribable content, the NLP service cannot produce a confidence score for filtered Files A and B. This shows that the data doesn't contain any speech discernible to AI speech recognition services.
\newline

Overall, our results demonstrate \names successfully preserved user privacy with nearly 100\% effectiveness through two evaluation methods.


\subsection{System Overhead}
The runtime for \names to predict airflow rate is between 2.9-3.4 ms, irrespective of smartphone models. This result implies that the computational overhead required for \names is relatively low. To better understand the impact of \names on battery life, we investigate the energy consumption of the entire \names stack, namely the application and OS-level APIs for sensing audio and context-awareness.  We compare this energy consumption against a baseline value consumed by the standard Android background processes and wireless radio, using the Android project's \textit{Battery Historian} tool \cite{googlebattery}.

\begin{table}[ht!]
\caption{\finalrev{\names energy consumption decreases with duty-cycling.}}
\label{tab:energyCon}
\vspace{-1em}
\scalebox{0.85}{
\begin{tabular}{|l|c|c|c|c|}\hline
\textbf{Application} & \textbf{Continuous} & \textbf{Duty cycle - 10 mins} & \textbf{Duty cycle - 15 mins} & \textbf{Duty cycle - 30 mins} \\\hline
\name & 6.08 mA/h & 0.61 mA/h & 0.42 mA/h & 0.21 mA/h \\
Android System Background & 6.60 mA/h & - & - & - \\\hline
\end{tabular}} 
\end{table}

One way to further reduce \name's energy consumption is implementing duty cycling, for example, sensing for one minute per $n$ interval. Table  \ref{tab:energyCon} summarizes our findings. In a typical workday setting of 8 hours, we can expect everyday users to utilize \names over the course of the day. Duty-cycling at 10 minutes and utilizing \names as an everyday application will only take up an additional 8.5\% \review{of} total consumption. With the battery capacities of modern smartphones today exceeding 2000 mA/h, we conclude that the background energy impact of \names would be negligible.
\newline


\section{Discussion and Future Work}
\label{sec:discussion}
Our study's objectives were to implement an audio-based sensing approach to measure the rate of airflow and better understand our model's efficacy under various indoor conditions and real-world interference. Here we discuss the implications of our findings.

\subsection{\finalrev{Fully Integrated Indoor Ventilation System}}
As clarified in Section \ref{sec:background}, the motivation of our work is set from the perspective of enabling proper ventilation in indoor environments. Proper ventilation essentially requires airflow to be monitored with other critical parameters such as CO\textsubscript{2}, humidity, temperature, and particulate matter. With our audio-sensing approach to predict the rate of airflow, we envision \names being a significant sub-system to a fully integrated, smart ventilation solution that provides healthy indoor air quality. 

\subsection{\finalrev{Empowering Users with Healthy Ventilation}}
\names can be utilized as a solution, which accumulates airflow data sensed from occupants' smartphones based on their indoor locations. For example, by coupling \names with an indoor-localization system, \names can be used to present maps that report poorly ventilated indoor spaces that users anonymously report. Similarly, occupants can track the status of healthy air ventilation, thus, empowering users with critical ventilation information to make better decisions about entering poorly ventilated buildings while assisting the organization stakeholder in maintaining standards.

\subsection{\finalrev{Robustness in Real-world Implementation}}

\subsubsection{\finalrev{Crowd Conditions}} We have shown how airflow rate can be predicted with audio signals. However, even though our experiments were tested on different vent outlets, buildings, and ambient noises, further studies will be required to determine the efficacy of \names in other large indoor settings. For example, the noise levels of crowded medium to large-sized dining rooms can typically range beyond 80 dB \cite{zemke2011little}. Crowded indoor conditions will likely not allow for silence period detection, a notable function of \name. Further, since ambient noise will be the most dominant signal, it is highly likely that MPS will not effectively identify the accurate airflow rate. However, such a scenario could benefit from our fixed audio-sensing platform, which requires implementing \names close to the air vent sources and maintaining the sounds of air from HVAC as the dominant signal.


\subsubsection{\finalrev{Handling Ambient Noises}} \review{Evidently, \names relies on silence detection and MPS mechanisms to accurately predict airflow rate. In situations where ambient noise is continuous, the hypothesis behind sensing for a silent period and MPS would break. However, our approach will be able to separate continuous noises lying in different audio spectrum (e.g., high-frequency: running a vacuum cleaner in the room\cite{jensen2010vacuum}). As our approach processes audio signals between 0-375 Hz, we foresee \names to produce erroneous predictions in cases with continuous low-frequency noises whose amplitude lies below the silent threshold. A possible workaround is to implement anomaly detection, which detects prediction results that are significantly different from the expected prediction series of the user. Our work continues to explore more sophisticated techniques to better handle these corner cases.}

\subsubsection{\finalrev{Calibration Procedure for Mobile App}} 
\review{Recall in Section \ref{sec:silentperiod}, the selection of threshold to detect silence period in our current implementation of \names is based on our dataset and smartphone devices. To improve system generalizability, \names must be calibrated for use in other indoor settings (e.g., shopping malls, airports, supermarkets) and users with different smartphone microphone specifications. At present, our dataset consists of indoor settings common to everyday office experience over two weeks. In a practical application, calibration in a new environment will require taking the smartphone close to the vent for few seconds to estimate the upper bound of vent noise (with some tolerance) for the silent threshold.}

\subsection{\finalrev{Extending Functionality of State-of-the-Art Airflow Sensors}}
Measuring the airflow rate is a first step to proving the feasibility of audio sensors as an alternative to airflow sensing. State-of-the-art airflow sensors such as pressure sensors and vane anemometers \cite{sundell2011ventilation,taylor2003measurement, mcwilliams2002review} can measure other characteristics related to air, including the direction of airflow. We believe it is technically feasible to estimate airflow direction using either multiple microphones or sensing airflow with a smartphone microphone at different orientations and positions. We can also use beam-forming to identify the location of the air source. These efforts remain as one direction of future work.


\section{Related Work}
Our focus here is to summarize existing literature on mobile sensing for building and health monitoring. Specifically, we highlight prior work that examined audio-sensing approaches to achieve these efforts.

\subsubsection{\finalrev{Mobile Sensing}}
Mobile sensing has been well established in the literature to provide vital information for air monitoring \cite{aram2012environment,thermo}. Some of these efforts include estimating ambient temperature through smartphone batteries \cite{breda2019hot}, determining human occupancy \cite{jiefan2018extracting} as a spatial characteristic to control HVAC ventilation, and estimating zone-based collaboration to calculate air exchange rate from temporal CO\textsubscript{2} concentration \cite{jiang2011maqs}. In the same vein, W-Air employs a low-cost gas sensor and wearable sensors to monitor indoor CO\textsubscript{2} concentrations \cite{maag2018w}. From reviewing prior work, one essential aspect of indoor air quality that we believe can benefit everyday users is recognizing how much air ventilation is required in the indoor space they are in. While this is not the central focus, determining airflow rate is the first step to realizing adequate indoor ventilation.

\subsubsection{\finalrev{Measuring Air with Microphone}}
Conceptually, there is prior work related to air sensing. Many of these efforts are geared towards utilizing the microphone sensor but for sensing human respiration~\cite{kulkas2009intelligent,shih2019breeze, larson2012spirosmart}. For example, Wang et al. developed a respiration monitoring system by sensing audio signal changes during respiration \cite{wang2018c}. Nam et al. detected nasal breath sound recordings from a microphone built-in smartphones to estimate respiration rate \cite{nam2015estimation}. Fundamentally, these works employ different audio-processing methods to extract frequency content from (breathing) audio signals. First, they identify respiratory cycle phases in the signals and then eliminate irrelevant motion noises to estimate physiological measures \cite{larson2012spirosmart,song2020spirosonic}. 

\subsubsection{\finalrev{Privacy in Audio Sensing}}
Despite the increasing recognition capability in audio sensing, working with audio as a primary data source often raises privacy concerns. In the context of an audio processing system for measuring airflow, the system should not learn anything about the user's speech. Many researchers have proposed techniques that use inaudible signals to fulfill a system's functionality\cite{tan2013sound}. For example, Sumeet et al. suggested randomly mutating raw sound frames and sub-sampling them to circumvent speech recovery \cite{kumar2015sound}. Other works include building speech classifiers and filtering these segments out when analyzing the core functionality, such as detecting coughs and breathing patterns \cite{ahmed2019mlung,liaqat2017method,larson2011accurate}. In the most recent study, Iravantchi et al. demonstrated how PrivacyMic takes advantage of inaudible frequencies to aid in acoustic activity recognition tasks \cite{iravantchi2021privacymic}. 
\newline

\noindent Overall, prior work further reinforces our decision to leverage an audio-sensing approach for measuring airflow. The richness in audio signals and methods to preserve privacy also indicate a promise to preserve user privacy when collecting audio signals.
\section{Conclusion}
Smartphones are increasingly adopted as a sensor or an extension to IoT devices that provide building monitoring capabilities. While much effort has focused on promoting energy-efficient and thermal comfort heating and air-conditioning and ventilation (HVAC) systems, ventilation has received much less attention despite its importance. This work proposed machine-learning algorithms to predict the state of an air vent (whether it is on or off) and the rate of air flowing through active vents based on an audio-sensing approach. We presented two techniques, silence period detection and Minimum Persistent Sensing, to enhance our machine learning methods to suppress human speech in sensed audio and reduce interference from ambient noise. \names is implemented as a fixed audio sensing platform on Arduino micro-controller and mobile audio sensing platform on smartphones. We validated our approach and demonstrated its efficacy and generalizability in controlled and real-world settings, accounting for different vent types, indoor environments, smartphone variations, and placements. \names as a fixed or mobile audio-sensing platform achieves over 90\% accuracy in predicting vent status and 0.96 MSE in predicting airflow rate when placed within 2.25 meters away from an air vent. Finally, our validation of the privacy-preserving pipeline from a user study and utilizing the Google Speech Recognition service found human speech inaudible and inconstructible. We discussed the promise and pitfalls of our work, complementing existing IoT devices to empower users with prioritizing healthier indoor ventilation when in indoor spaces. Our application and datasets are available here: [\url{https://github.com/umassos/FlowSense}].

\section*{Acknowledgments} 
We thank the anonymous reviewers for their suggestions for improving the paper. This research was supported in part by NSF grants 2021693, 2020888, 1836752, and  US Army contract W911NF-17-2-0196. Any opinions, findings, conclusions, or recommendations expressed in this paper are those of the authors and do not necessarily reflect the views of the funding agencies.


\bibliographystyle{unsrt}
\bibliography{main}


\end{document}